\documentclass[prb,noshowpacs,footbib,superscriptaddress,showpacs,twocolumn]{revtex4}

\usepackage{graphicx}
\usepackage{amssymb,amsmath}
\usepackage[colorlinks=true,linktoc=page,linkcolor=blue,citecolor=magenta]{hyperref}

\usepackage{alphalph}
\usepackage{braket}

\usepackage{natbib}
\newcommand{\kk}{{\mathbf{k}}}
\newcommand{\GG}{{\mathbf{G}}}

\newcommand{\dd}{{\mathbf{d}}}
\newcommand{\pp}{{\mathbf{p}}}
\newcommand{\im}{{\mathrm{i}}}
\newcommand{\oneone}{{\mathbb{I}}}
\renewcommand{\Re}{\mathrm{Re\,}}
\renewcommand{\Im}{\mathrm{Im\,}}

\begin{document}

\title{Topological electronic properties of silicon}

\author{A. Shtyk}
\affiliation{Department of Physics, Harvard University, Cambridge, MA 02138, USA}
\author{C. Chamon}
\affiliation{Department of Physics, Boston University, Boston, MA, 02215, USA}



\begin{abstract}
The central role that materials play in human history is exemplified
by the three-age division of prehistory into the stone, bronze, and
iron ages. References to our present time as the information age or
silicon age epitomizes the important role that this semiconducting
material came to play in the development of computers and devices that
permeate our daily lives. Here we show that the electronic states in
silicon have nontrivial topological structures that are captured by a
network of Berry flux lines that link at points of high symmetry in
the Brillouin zone. This complex network has ice-nodal points where
fluxes satisfy ice rules, making silicon a ``nodal-chain
insulator''. Fixing the longitudinal momentum parallel to such flux
lines yields a two-dimensional Dirac Hamiltonian for the transverse
degrees of freedom. This complex Berry-flux network implies a
topologically stable two-fold degeneracy along the X-W direction in
all of silicon bands, a fact that is supported by crystal symmetry
arguments as well as direct inspection of the vast literature on
silicon band structures. Similarly to nodal-chain semimetals, we find
drumhead-like states in the regions that are delimited by the
projections of the bulk Berry flux network on the surface Brillouin
zone.

\end{abstract}
\maketitle


\section{Introduction}

The experimental discovery of the integer quantum Hall
effect~\cite{vonKlitzing80}, where the Hall resistance is quantized to
the extraordinary precision of one part in a billion, led to the new
standard of resistance for the international system of units. A degree
of precision such as this has its roots in a fruitful confluence of
physics and mathematics, which ties the Hall resistance to a
topological quantity. In the case of the integer Hall effect, this
quantity is the first Chern number associated to each filled Landau
level~\cite{TKNN82}. The integer Hall effect was the first example of
a system with topological electronic properties; the number of systems
in which topology plays a prominent role has grown explosively in the
recent past, fueled by the discovery of a new class of topological
band insulators occurring in semiconductors with strong
spin-orbit coupling, in which gapless surface states
exist~\cite{Kane05a,Kane05b,Fu-Kane06,Bernevig06a,Bernevig06b,Konig07,Fu07,Moore07,Roy09a,Roy09b,Hsieh08,Hsieh09,Hsieh09b,Chen09}
(For reviews, see Refs.~\onlinecite {Hasan-Kane10,Qi-Zhang11}.) After the
discovery of topological insulators, many examples of topological
semimetals were identified, such as Weyl
metals~\cite{BurkovHookBalents,BurkovBalents,Wan11,Neupane2014,Liu14,Huang2015,Lv15}
and systems with Weyl nodal lines~\cite{BurkovHookBalents,Yu15}, and
nodal chains~\cite{bzdusek_etal_2016}.

Spin-orbit interactions play a prominent role in the topological
insulators and Weyl systems, but spin-orbit coupling is not central to
the understanding of the basic electronic properties of
silicon. Indeed, in the standard classification of topological
insulators, silicon is not classified~\cite{Schnyder08} as one with
protected Dirac surface states. Therefore, silicon thus far has sat on
the sidelines. Here we show that the sublattice structure of crystals
such as silicon is responsible for a network of Berry flux lines in
the Brillouin zone that link at points of high symmetry in momentum
space. This Berry flux network is topologically stable, obeys ice
rules (two in, two out) at the X points, and is responsible for
topological protection of degeneracies along the X-W direction. The
nontrivial topological structure of the Berry flux network in silicon
shares the same physical origin as the Berry flux in graphene: the
fact that there are two atoms in the unit cell gives rise to a spinor
structure with associated Berry phases. The existence of the Berry
flux network opens a novel ``topological knob'' to manipulate
electrons in silicon, especially in light of progress made in the past
decade from studying the effects of Berry phases in the electronic
properties in graphene.

\section{Berry flux network}

We uncover the Berry flux network in two steps. First we identify the
spinor structure and the Berry fluxes within a general tight-binding
approximation (valid for an arbitrary number of orbitals). Second we
argue based on topological and symmetry arguments that the Berry flux
network is robust and remains beyond the tight-binding description of
silicon.

A tight-binding Hamiltonian for a system on a bipartite lattice
comprised of sublattices $A$ and $B$ can be represented as
\begin{equation}
  H(\kk)
  =
  \begin{bmatrix}
    H_{AA}(\kk)&H_{AB}(\kk)\\
    H_{BA}(\kk)&H_{BB}(\kk)
  \end{bmatrix}
  \;,
\end{equation}
where the blocks $H_{AA}$ and $H_{BB}$ contain hoppings between sites
in the same sublattice, and the blocks $H^{\;}_{AB}$ and
$H^{\;}_{BA}=H^\dagger_{AB}$ contain hoppings between sites is
different sublattices. The size of these blocks depend on the number
of orbitals included in the tight-binding model. For example, in
graphene the blocks are $1\times 1$ if one considers only the
$\pi$-orbital, and in silicon the blocks are $4\times 4$ if one takes
account of only the $s,p_x,p_y,p_z$ orbitals (or $5\times 5$ in the
$sp^3s^*$ model~\cite{VOGL1983}, with the inclusion of the $s^*$
orbital). The diagonal blocks are periodic in
$\kk$-space: $H_{AA}(\kk)=H_{AA}(\kk+\GG_i)$ and
$H_{BB}(\kk)=H_{BB}(\kk+\GG_i)$, where $\GG_i$ ($i=1,\dots,d$) is a
reciprocal lattice basis vector ($d=2$ for graphene, $d=3$ for
silicon). The matrix elements of the off-diagonal blocks are written
as
\begin{equation}
  \label{eq:off-diagonal_H}
  \left[H^{\;}_{AB}(\kk)\right]_{\alpha\beta}
  =
  -
  \sum_{\mu}\;
  t_{\alpha\beta}(\dd_\mu)\;e^{\im\kk\cdot\dd_\mu}
       \;,
\end{equation}
where the vectors $\dd_\mu$ connect the atoms from $A$ to $B$, and the
$t_{\alpha\beta}(\dd_\mu)$ contain the overlap of the orbitals
$\alpha$ and $\beta$ separated by $\dd_\mu$. Because the vectors
$\dd_\mu$ are not Bravais lattice vectors, $\dd_\mu\cdot G_i$ is
generically not a multiple of $2\pi$, and therefore the off-diagonal
blocks $H_{AB}$ and $H_{BA}$ are {\it not} periodic in $\kk$-space. An
attempt to gauge out these non-periodicities comes with the price of
adding singularities (branch cuts) to the phase of the wave function
in momentum space.

\begin{figure}[t]
	\center{\includegraphics[width=.6\linewidth]{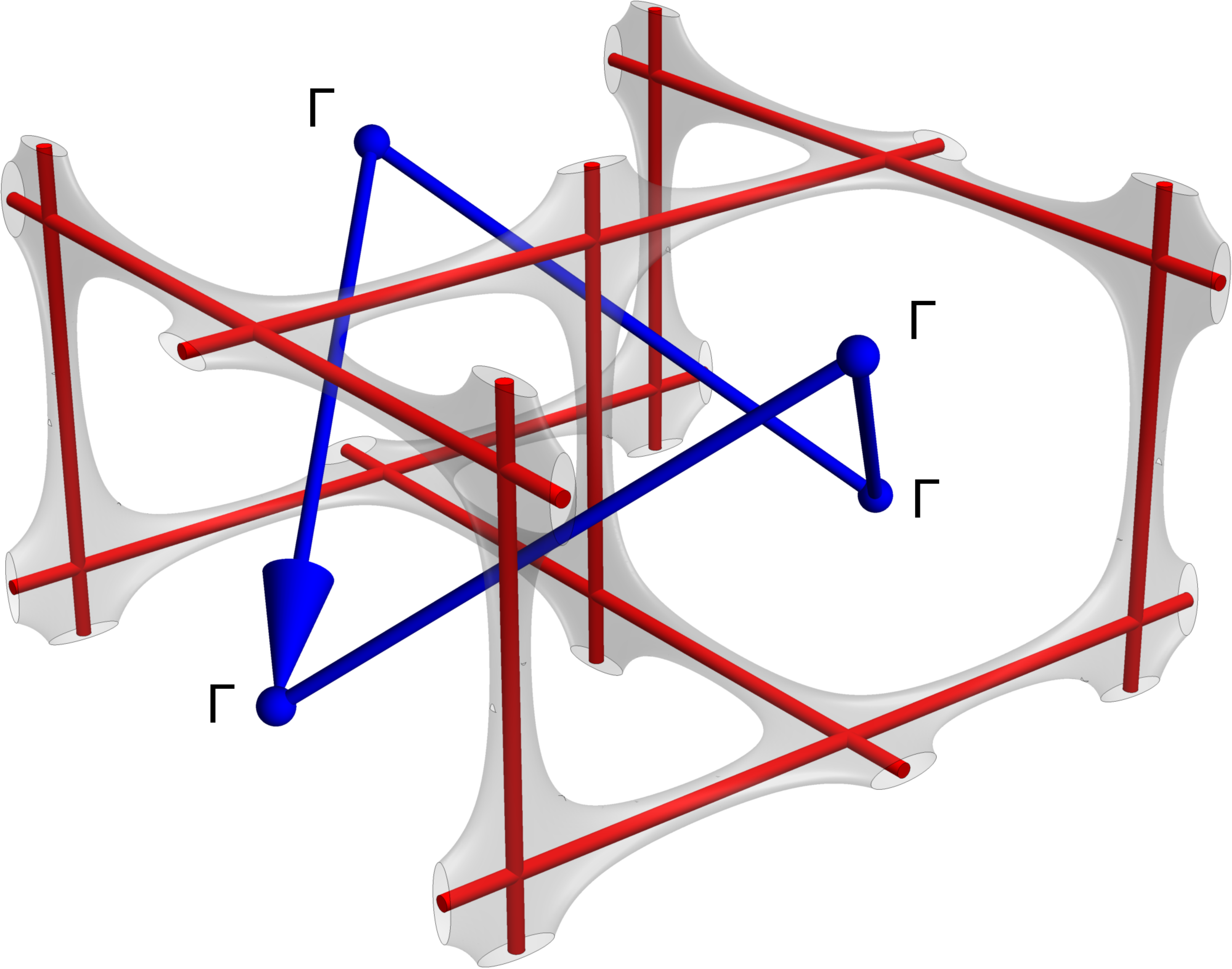}}
	\caption{{\bf A loop in $\kk$-space encircling a single $\pi$-flux line.} A Berry flux network in $\kk$-space is schematically shown in red color with the grey surface guiding the visualization of the flux flow. Blue color represents a $\mathbf{0}$--$\GG_1$--$(\GG_1+\GG_2)$--$(\GG_1+\GG_2+\GG_3)$--$\mathbf{0}$ walk that we use in the main text to argue the existence of the Berry $\pi$-flux /Dirac line piercing this loop.}
	\label{fig:G_walk}
\end{figure}

In graphene the vectors $\dd_\mu$, $\mu =0,1,2$, point to the vertices
of a triangle, while in silicon the $\dd_\mu$, $\mu =0,1,2,3$, point
to the vertices of a tetrahedron. In these lattices (see Supplementary
Online Information), it follows that $H_{AB}(\kk+\GG_i)= e^{\im
  \Phi(\GG_i)}\;H_{AB}(\kk)$, where $\Phi(\GG_i)=2\pi/N$, with $N=3$
and $N=4$ for the graphene and silicon lattices, respectively. The
Hamiltonian $H(\kk)$ is not periodic in $\kk$-space; however, it is
periodic {\it up to} a unitary transformation that rotates the
amplitudes on the two sublattices by opposite phases:
\begin{equation}
  \label{eq:H-gauge}
  H(\kk+\GG_i)
  =
  U(\GG_i)\;H(\kk)\;U^\dagger(\GG_i)
  \;,
\end{equation}
with
\begin{equation}
U(\GG_i)
=
e^{\;\im\frac{1}{2}\Phi(\GG_i)\;\sigma_z},
\end{equation}
and $\sigma_z$ a Pauli matrix acting on the $A/B$ sublattice grading.
It follows from Eq.~(\ref{eq:H-gauge}) that the eigenenergies
$\epsilon(\kk)=\epsilon(\kk+\GG_i)$, as expected. However, the spinor
structure and the lack of periodicity of the matrix Hamiltonian
$H(\kk)$ (not just its eigenvalues) is what leads to the Berry
$\pi$-vortices at the $K$ points in graphene, and the Berry $\pi$-flux
network that we uncover in this work. (In the Supplementary Online
Information we illustrate how the generic framework above recovers the
familiar results in graphene as a warm up for the calculations in
silicon.)

Consider the walk in $\kk$-space that visits, in order, the points
$\kk,\kk+\GG_1,\kk+\GG_1+\GG_2, \kk+\GG_1+\GG_2+\GG_3$ and back to
$\kk$. This walk passes through 4 of the 6 edges of the tetrahedron
formed by the four points in $\kk$-space, closing a loop. We choose
the initial point not to be one of high symmetry, to avoid that the
edges pass through band crossings, thus avoiding degeneracies along
the walk. For example, one may choose to start close to but not at the
$\Gamma$ point, say at $\kk=(\delta_x,\delta_y,\delta_z)$, with
infinitesimal $\delta_{x,y,z}$. At the end of the walk, the
Hamiltonian returns to $H(\kk)$, but the eigenvector is rotated by the
sequence of unitary operations
\begin{equation}
  \label{eq:4_Us}
  U(-\GG_1-\GG_2-\GG_3)\;U(\GG_3)\;U(\GG_2)\;U(\GG_1)
  =
  e^{\;\im\pi\;\sigma_z}=-\oneone
  \;,
\end{equation}
which amounts to a rotation by $\pi$~\,\cite{Berry_gauge_choice}.  This
geometric phase implies the existence of $\pi$-flux lines in
$\kk$-space, which pierce the loop we described above. An example of a
$\pi$-flux network that threads the 4-edged loop in $\kk$-space is
shown in Fig.~\ref{fig:G_walk}, which, as we show below, corresponds
to the cases of the lowest conduction and valence bands in silicon.

These singular flux lines are stable, and cannot be removed by small
deformations. Silicon is inversion symmetric, which implies that the
Berry curvature ${\cal B}(\kk)={\cal B}(-\kk)$; in the absence of
time-reversal breaking perturbations, ${\cal B}(\kk)=-{\cal
  B}(-\kk)$. These two symmetries, together, imply that the Berry
curvature vanishes everywhere with the possible exception of singular
lines carrying flux multiple of $\pi$, like those we identified
above. These two symmetries thus ensure that the $\pi$-flux cannot
spread over a finite region, and thus remains singular and contained
within a network of flux lines circulating around the Brillouin
zone. The number of orbitals in the description of the system does not
alter our conclusions based on the topological constraints imposed by
Eq.~(\ref{eq:4_Us}).

\begin{figure}[b]
	\begin{minipage}[h]{.49\linewidth}
		\center{\includegraphics[width=1\linewidth]{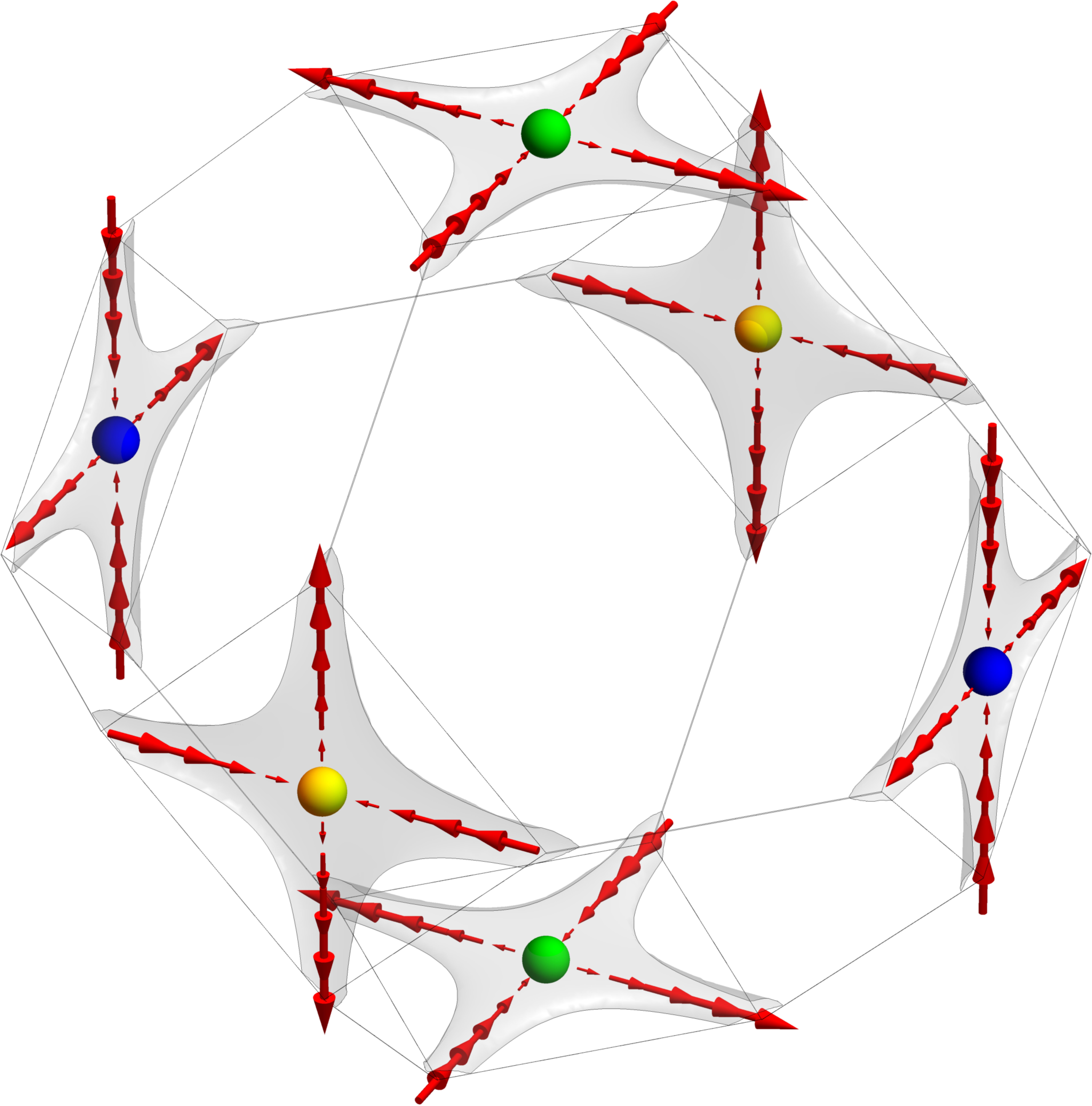}}
	\end{minipage}
	\hfill
	\begin{minipage}[h]{.49\linewidth}
		\center{\includegraphics[width=1\linewidth]{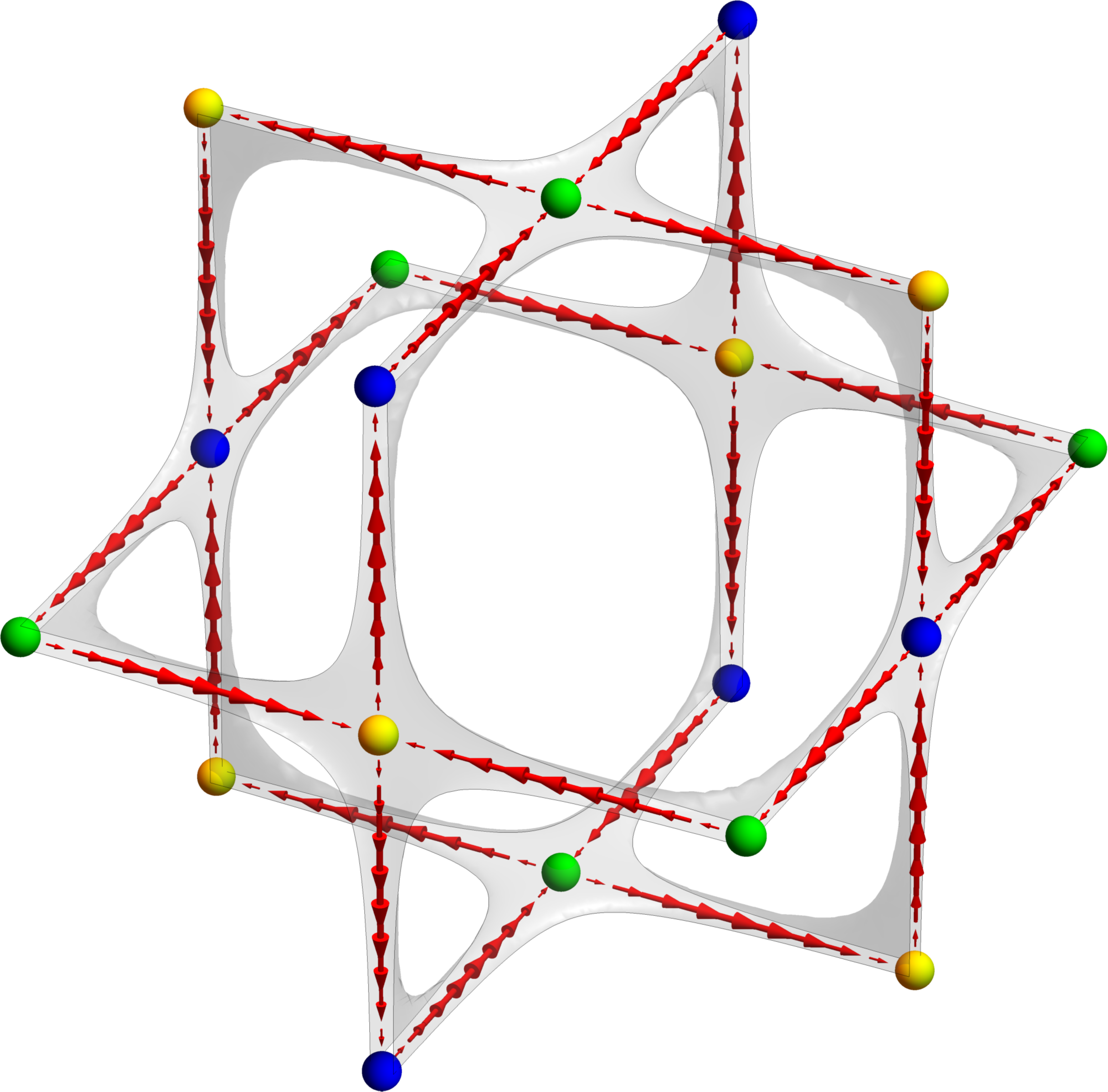}}
	\end{minipage}
	\caption{{\bf Berry curvature field.} \textit{Left:} a single first Brillouin zone. \textit{Right:} a doubled Brillouin zone. The colored spheres represent three inequivalent X points. The grey surface guides the visualization of the flux flow and the ice-rule links at the X points.}
	\label{fig:wire_frame}
\end{figure}

\begin{figure}[b]
	\begin{minipage}[h]{0.63\linewidth}
		\center{\includegraphics[width=1\linewidth]{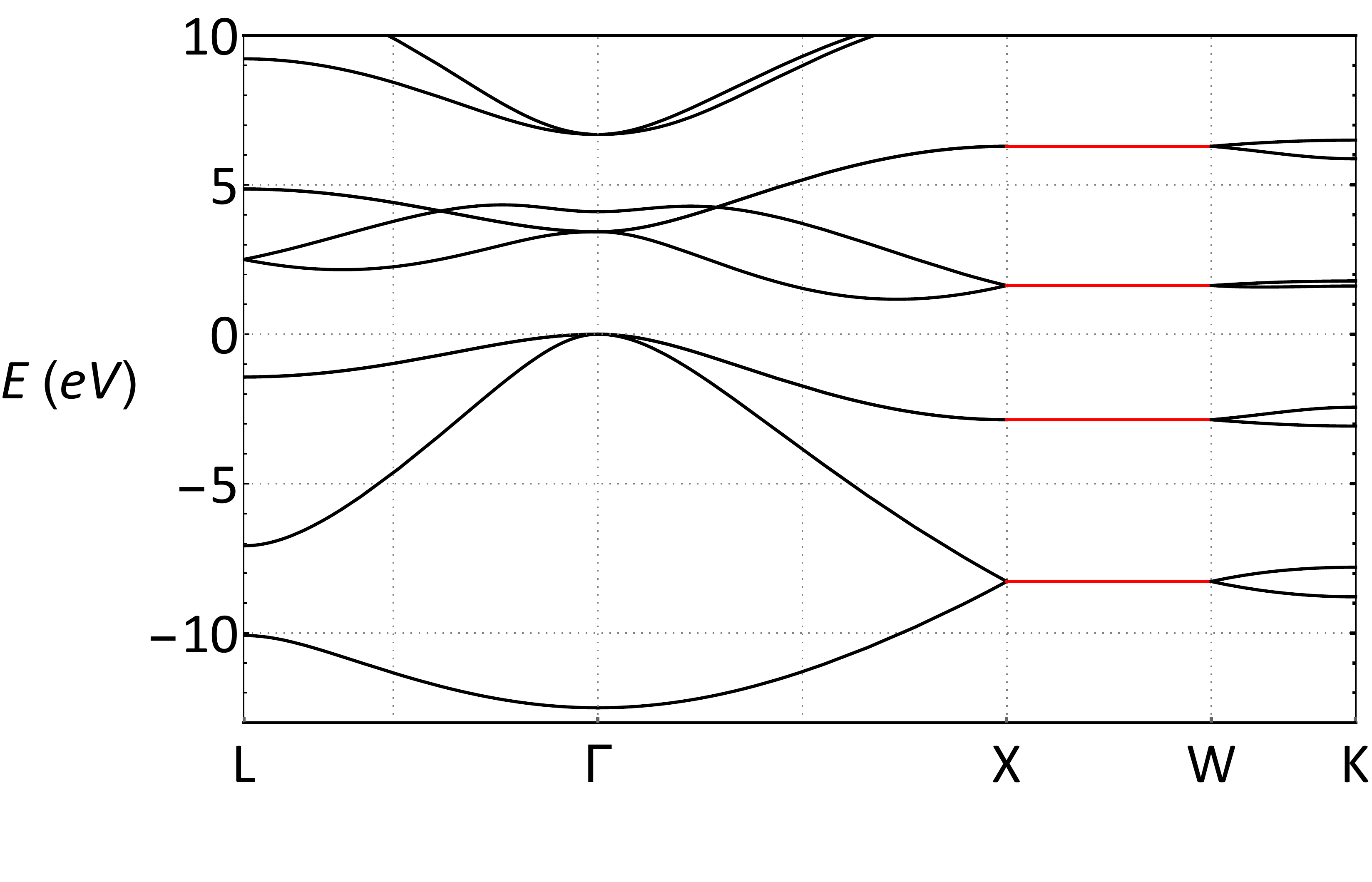}}
	\end{minipage}
	\hfill
	\begin{minipage}[h]{0.35\linewidth}
		\center{\includegraphics[width=1\linewidth]{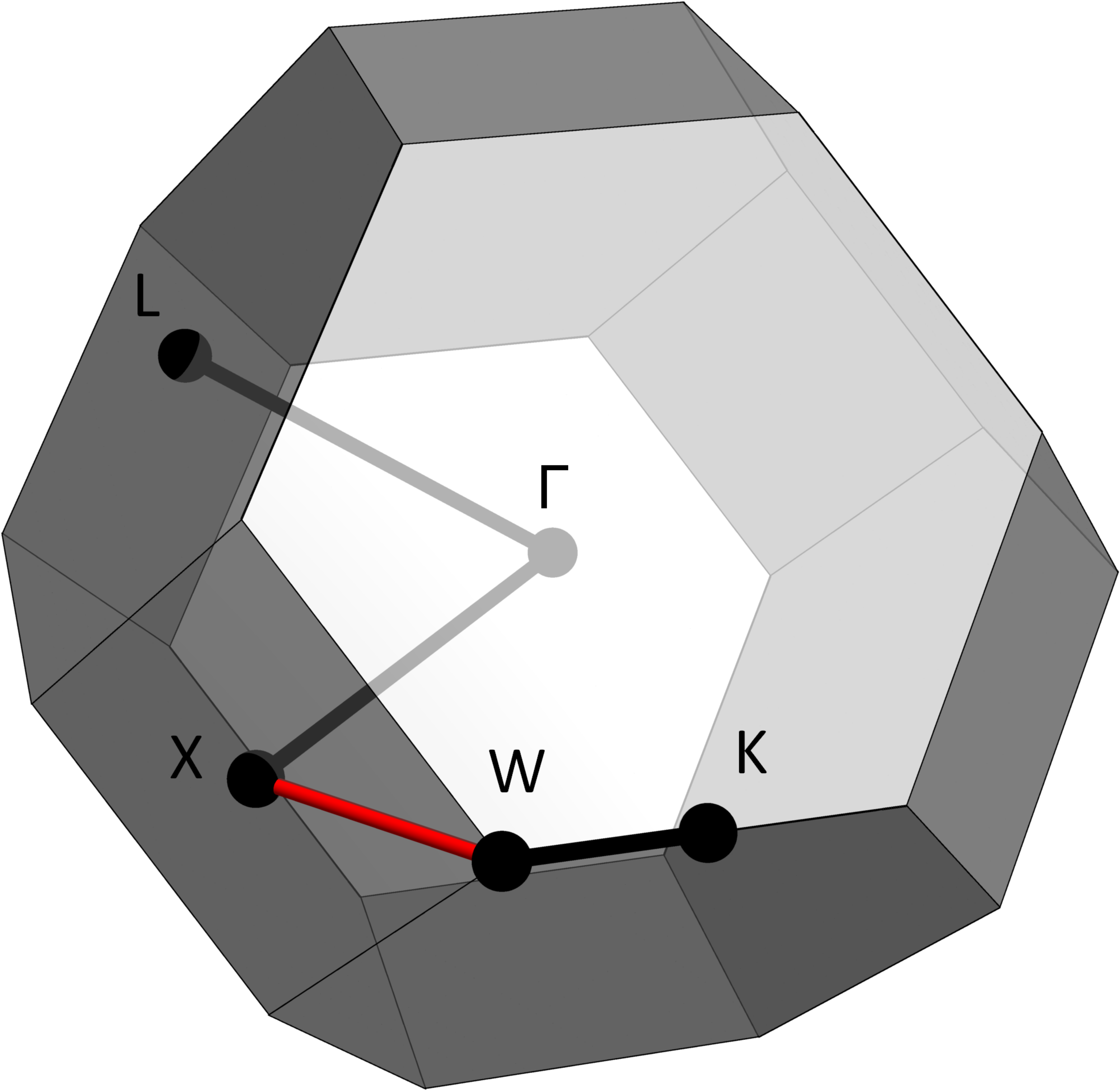}}
	\end{minipage}
	\caption{{\bf Electronic band structure of silicon within the $sp^3s^*$ model.} \textit{Left:} all bands exhibit a two-fold degeneracy along the X-W line (highlighted in red). This degeneracy may be used as evidence for the existence of the Dirac lines along the X-W direction. \textit{Right:} first Brillouin zone with points of high symmetry.}
	\label{fig:band_structure}
\end{figure}

To visualize the network of fluxes in silicon, we consider explicitly
the 5-orbital $sp^3s^*$ nearest-neighbor tight-binding
model~\cite{VOGL1983}. This model captures essential features of
silicon's band structure; in particular, it reproduces the conduction
band minimum along the $\Delta$ line connecting the $\Gamma$ and X
points in the Brillouin zone. In this approximation, sites within the
same sublattice are not connected, so that diagonal blocks simply
contain the on-site potential energies $H_{AA/BB}=\text{diag}(E_s,
E_p, E_p, E_p, E_{s^*})$, while the inter-sublattice hoppings contain
the non-trivial momentum dependence:
\begin{widetext}
\begin{equation}
H_{AB}(\kk)=
\begin{bmatrix}
\phantom{-}V_{ss}\;g_0(\kk) & V_{sp}\;g_1(\kk) & V_{sp}\;g_2(\kk) & V_{sp}\;g_3(\kk) & 0
\\
-V_{sp}\;g_1(\kk) & V_{xx}\;g_0(\kk) & V_{xy}\;g_3(\kk) & V_{xy}\;g_2(\kk) & -V_{s^*p}\;g_1(\kk)
\\
-V_{sp}\;g_2(\kk) & V_{xy}\;g_3(\kk) & V_{xx}\;g_0(\kk) & V_{xy}\;g_1(\kk) & -V_{s^*p}\;g_2(\kk)
\\
-V_{sp}\;g_3(\kk) & V_{xy}\;g_2(\kk) & V_{xy}\;g_1(\kk) & V_{xx}\;g_0(\kk) & -V_{s^*p}\;g_3(\kk)
\\
0 & V_{s^*p}\;g_1(\kk) & V_{s^*p}\;g_2(\kk) & V_{s^*p}\;g_3(\kk) & 0
\end{bmatrix}
\end{equation}
\end{widetext}
where momentum functions
\begin{equation}
\begin{split}
g_0(\kk)=&\frac{1}{4}\left(
e^{i\dd_0\cdot\kk}
+e^{i\dd_1\cdot\kk}
+e^{i\dd_2\cdot\kk}
+e^{i\dd_3\cdot\kk}
\right),
\\
g_1(\kk)=&\frac{1}{4}\left(
e^{i\dd_0\cdot\kk}
+e^{i\dd_1\cdot\kk}
-e^{i\dd_2\cdot\kk}
-e^{i\dd_3\cdot\kk}
\right),
\\	
g_2(\kk)=&\frac{1}{4}\left(
e^{i\dd_0\cdot\kk}
-e^{i\dd_1\cdot\kk}
+e^{i\dd_2\cdot\kk}
-e^{i\dd_3\cdot\kk}
\right),
\\
g_3(\kk)=&\frac{1}{4}\left(
e^{i\dd_0\cdot\kk}
-e^{i\dd_1\cdot\kk}-
e^{i\dd_2\cdot\kk}
+e^{i\dd_3\cdot\kk}
\right),
\end{split}
\end{equation}
and \; $\dd_0=\frac{a}{4}\, (1 \, 1\, 1)$, $\dd_1=\frac{a}{4}\, (1 \,
\bar{1} \, \bar{1})$, $\dd_2=\frac{a}{4}\, (\bar{1} \, 1 \, \bar{1})$,
and $\dd_3=\frac{a}{4}\, (\bar{1} \, \bar{1} \, 1)$, with
$a=5.4310$\AA. The interaction parameters in the Hamiltonian are (in
eV) $E_s=-4.20, E_p=1.72, E_{s^*}=6.69, V_{ss}=-8.30, V_{sp}=5.73,
V_{s^*p}=5.38, V_{xx}=1.72, V_{xy}=4.58$~\cite{VOGL1983}. The
resulting band structure is shown in Fig.~\ref{fig:band_structure}.

This spectrum has an intricate set of lines along which the spectrum
is twofold degenerate, in particular along the Z line connecting the X
and W points, which is consistent with the $O^7_h$ crystal symmetry of
silicon~\cite{Slater_book}. We argue that these degeneracies,
associated with the Berry flux $\pi$ identified above, correspond to
Dirac lines: fixing the longitudinal momentum along the line yields a
two-dimensional Dirac Hamiltonian for the transverse degrees of
freedom. While the dispersionless nature of these lines along their
longitudinal direction is an artifact of the nearest-neighbor
tight-binding approximation, the two-fold degeneracy and the Berry
$\pi$-flux that travels along these lines are robust. We remark that
distorting the hoppings would not remove these lines of degeneracy
because the $\pi$-fluxes are topologically stable; hence topology
ensures that there should be lines of degeneracy in silicon even if
rotational symmetry is broken but sublattice symmetry is not. (An
analogous reasoning holds in graphene, where the nodes are stable even
if the hopping matrix elements to the three neighbors are close but
unequal.) We find that the lowest conduction and valence bands exhibit
the simplest pattern of these Berry flux lines, which reduce to a
cage-like net of Dirac lines connecting inequivalent X points (going
along the Z line through both X and W points), as is shown on the
Fig.~\ref{fig:wire_frame}. The Dirac lines meet at the X point,
forming an ice-nodal point (Fig.~\ref{fig:X_point}).

\section{Effective Hamiltonian near the X point of the lowest conduction band}

\begin{figure}
	\center{\includegraphics[width=0.5\linewidth]{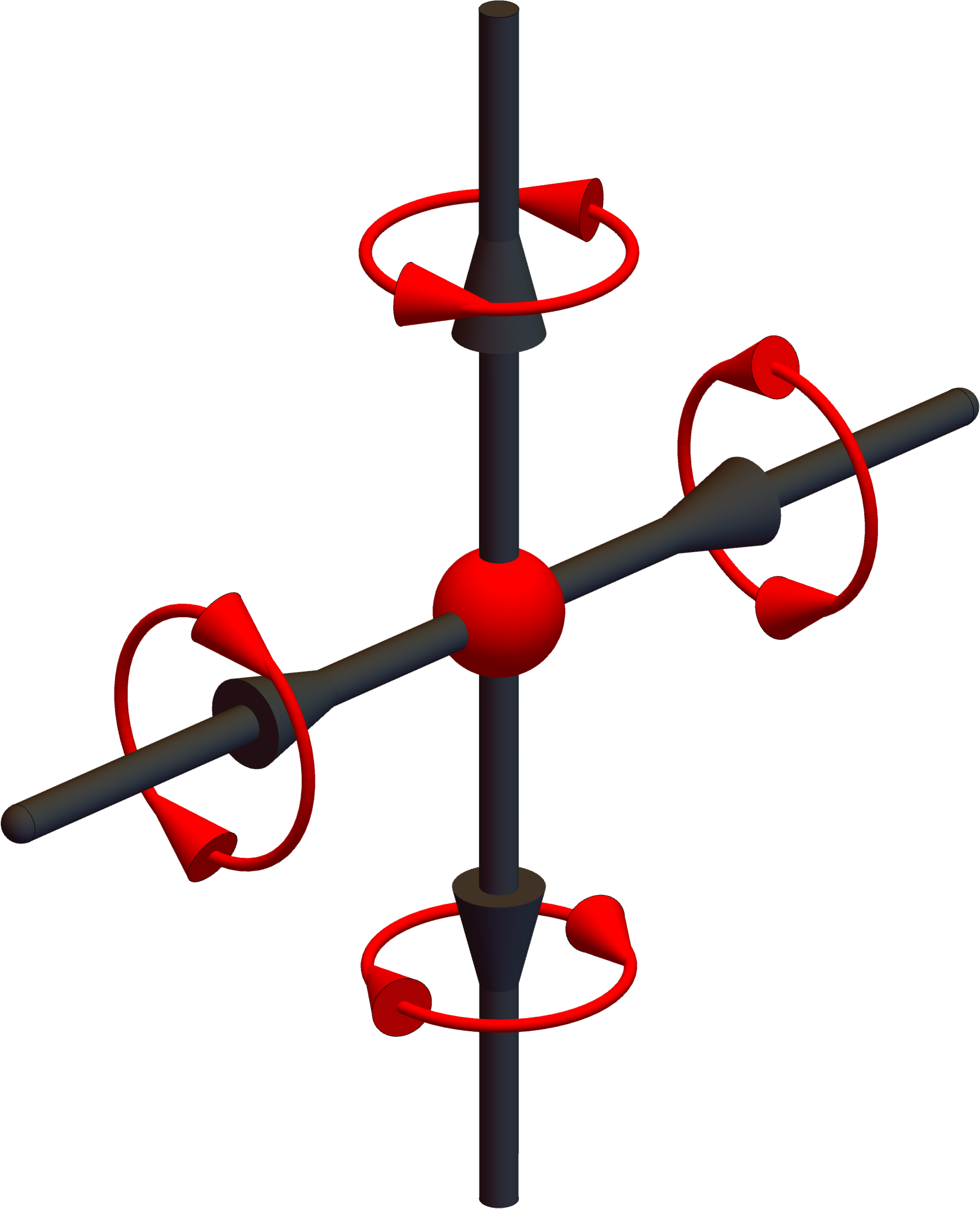}} 
	\caption{{\bf Berry flux network near the X point.} The figure shows Dirac lines linking at the X point according to the ice rule 2-in and 2-out. Red loops show the winding of the phase around the lines to visualize the ice rule.}
	\label{fig:X_point}
\end{figure}

The effective two-band Hamiltonian in the vicinity of the ice-nodal
point of the first conduction band is obtained by expanding
$\kk=(2\pi,0,0)+\pp$, yielding (see Supplementary Online Information
for details)
\begin{equation}
H_{X}=\varepsilon_0(\pp)\,\hat{\sigma}_0+v_x\,p_x\,\hat{\sigma}_1+\kappa_{yz}\,p_y\,p_z\,\hat{\sigma}_2, 
\end{equation}
where $v_x=0.51$, $\kappa_{yz}=0.18$, $\varepsilon_0(\vec{p})=1.63 +
0.11\,p_x^2$, and $\hat{\sigma}_i$ are Pauli matrices. (Energies are
measured in eV and momentum in units of $1/a$.) This expression
explicitly shows the ice-nodal nature of the X point. For example, in
the vicinity of the points $2\pi\,(1, 0, \pm0.1)$ the Hamiltonian takes the
form
\begin{equation}
H_{X\pm\delta}=\tilde{v}_x\,p_x\,\hat{\sigma}_1 \pm \tilde{v}_y\,p_y\,\hat{\sigma}_2,
\end{equation}
where $\tilde{v}_x=0.51$ and $\tilde{v}_y=0.02$ are electron
velocities at the $2\pi(1, 0, \pm0.1)$ points. This Hamiltonian structure
indicates two Dirac lines with opposite chiralities approaching the X
point along the $z$-axis from opposite directions. Similarly, there
are two more such lines along the $y$-axis (see Fig.~\ref{fig:X_point}).

It turns out that all crucial topological properties of the conduction
band can be studied within a simple toy model, with a single orbital
per site and nearest-neighbor tight-binding Hamiltonian
\begin{equation}
\label{eq:toy}
\begin{split}
H_{\text{toy}}(\kk)
=&
t_{nn}\begin{bmatrix}
0 & g_0(\kk)
\\
g_0^*(\kk) & 0
\end{bmatrix}
\\
=&
t_{nn}\,
\left(\hat{\sigma}_1\;\Re g_0(\kk)+\hat{\sigma}_2\;\Im g_0(\kk)\right).
\end{split}
\end{equation}
Hopping matrix elements to farther neighbors can be included,
especially between the sites of the same sublattice, to reproduce
features of silicon's band structure, such as a conduction band
minimum along the $\Delta$ line. Yet, the topological features are
captured by the off-diagonal terms alone: the Hamiltonian is
degenerate along the $g_0(\kk)=0$ manifold, yielding essentially the
same cage-like net of Dirac lines connecting at X points, just as in
the full $sp^3s^*$ model for silicon.

The real and imaginary parts of $g_0(\kk)$ are
\begin{align}
\Re g_0(\kk)=\cos\frac{k_x}{4}\cos\frac{k_y}{4}\cos\frac{k_z}{4},
\nonumber\\
\Im g_0(\kk)=-\sin\frac{k_x}{4}\sin\frac{k_y}{4}\sin\frac{k_z}{4},
\end{align}
from which we identify
\begin{align}
& (2\pi,0,s),\; (2\pi,s,0),\; (0,2\pi,s),\; 
  (s,2\pi,0),  \nonumber\\
  &\; (0,s,2\pi),\; 
 \text{and}\;
  (s,0,2\pi)\;, 
 \; \text{for}\; s\in [0,2\pi),
\end{align}
as the set of nodal lines, crossing at the three ice-nodal X points,
\begin{equation}
(2\pi,0,0),\quad(0,2\pi,0),\quad(0,0,2\pi).
\end{equation}
Expanding around one of the ice-nodal points, $\kk=(2\pi,0,0)+\pp$, we
obtain
\begin{equation}
\begin{split}
\Re g_0(\kk)=&-\sin\frac{p_x}{4}\cos\frac{p_y}{4}\cos\frac{p_z}{4}\simeq-\frac{1}{4}p_x,
\\
\Im g_0(\kk)=&-\cos\frac{p_x}{4}\sin\frac{p_y}{4}\sin\frac{p_z}{4}\simeq-\frac{p_yp_z}{16},
\end{split}
\end{equation}
reproducing the same structure of the X-point effective Hamiltonian
derived using the $sp^3s^*$ model:
\begin{equation}
H_{\text{toy, X}}=-V_{AB}
\left(\frac{p_x}{4}\hat{\sigma}_1+\frac{p_y\,p_z}{16}\hat{\sigma}_2\right)
\;.
\end{equation}
We note that perturbing this toy model Hamiltonian with a
$\hat\sigma_3$ term gaps out the Dirac lines (it breaks sublattice
symmetry), while a $\hat\sigma_2$ perturbation separates the Dirac
lines in different ways depending on the sign of the $\hat\sigma_2$ term,
indicating the critical character of the ice-nodal point.

Finally, we gather all results above on the ice-nodal points and
combine them with other information inferred from results on the band
structure of silicon that is obtained from methods other than
tight-binding. We condense this combined information into an effective
Hamiltonian near the $(2\pi,0,0)$ nodal point:
\begin{equation}
\label{eq:H_eff}
H_{\text{eff, X}}=
\frac{p_x^2}{2m_\ell}+\frac{p_y^2}{2m_t}+\frac{p_z^2}{2m_t} +
v_x\;p_x\;\hat{\sigma}_1+\kappa\;p_y\,p_z\;\hat{\sigma}_2 \;,
\end{equation}
where $m_\ell=0.98\,m_e$ and $m_t=0.19\,m_e$ coincide with the masses
from the standard low-energy description of the conduction band (with
$m_e$ the electron mass); the velocity
$v_x=0.15\;({2\pi}/{a})({\hbar}/{m_\ell})$ is estimated from the
location of the conduction band minimum; and $\kappa$ follows from the
dispersion along the X-U direction. This effective Hamiltonian, we
claim, captures the topological properties and hence gives a more
accurate description of the low energy states in silicon's conduction
band than simply expanding the dispersion to quadratic order near the
band minimum. That the minimum occurs close to the X point is captured
by the interplay between the Dirac dispersion along the $x$-direction
and the parabolic term, which places the minimum close to the X point
(notice that the energy difference between the minimum and the X point
is only 0.08eV~\cite{ioffe}).

\section{Surface states and bulk-boundary correspondence}

Probably the most important theoretical insight in the field of
topological insulators is the fact that non-trivial bulk topology
leads to the existence of robust surface states. Recently this idea
was extended to topological semimetals, where the Fermi surface consists
of a nodal loop. In such materials surface states exist only in parts
of the surface Brillouin zone that are determined by projecting the
nodal loop from the bulk Brillouin zone onto the surface Brillouin
zone. Here we argue that the same bulk-boundary correspondence applies
to silicon and the existence of the Berry flux ``wire frame'' leads to
the existence of non-trivial topological drumhead surface states.

\begin{figure}[t]
	\center{\includegraphics[width=.8\linewidth]{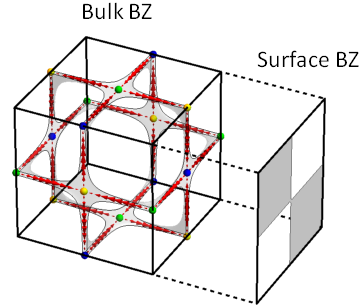}}
	\caption{{\bf Bulk-boundary correspondence within the toy model.} (Doubled) bulk and surface Brillouin zones are topologically related. Bulk wire frame projected on the surface Brillouin zone breaks it into quadrants. Surface drumhead state exists either in first and third or second and fourth, depending whether the crystal surface terminates on A or B sublattice }
	\label{fig:bulk_boundary}
\end{figure}

The two-band toy model Hamiltonian (\ref{eq:toy}) is of an archetypal
semimetallic and the manifold $g_0(\kk)=0$ defines a nodal
loop. Moreover, this Hamiltonian describes a nodal chain semimetal, a
topologically critical system with crossing nodal lines yielding
ice-nodal X
points\cite{bzdusek_etal_2016,yan2018experimental}. According to the
intuition from semimetallic
systems\cite{atala2013,rhim2017bulk,van2016topological,naumov2016topological},
the projection of the bulk wire frame onto the surface Brillouin zone
breaks it into segments. These segments can be colored with two
colors, for example red and blue, in such way that adjacent parts are
always of the opposite color. Surface drumhead states then exist in
all segments colored either in blue or in red, depending whether the
crystal surface is terminated at A or B sublattice. In the case of
silicon that we consider, the projection of the wire frame breaks down
surface Brillouin zone into quadrants. Surface states exist either in
first and third or second and fourth quadrants, see
Fig.~\ref{fig:bulk_boundary}.

\begin{figure}[b!]
	\begin{minipage}[h]{0.49\linewidth}
		\center{\includegraphics[width=1\linewidth]{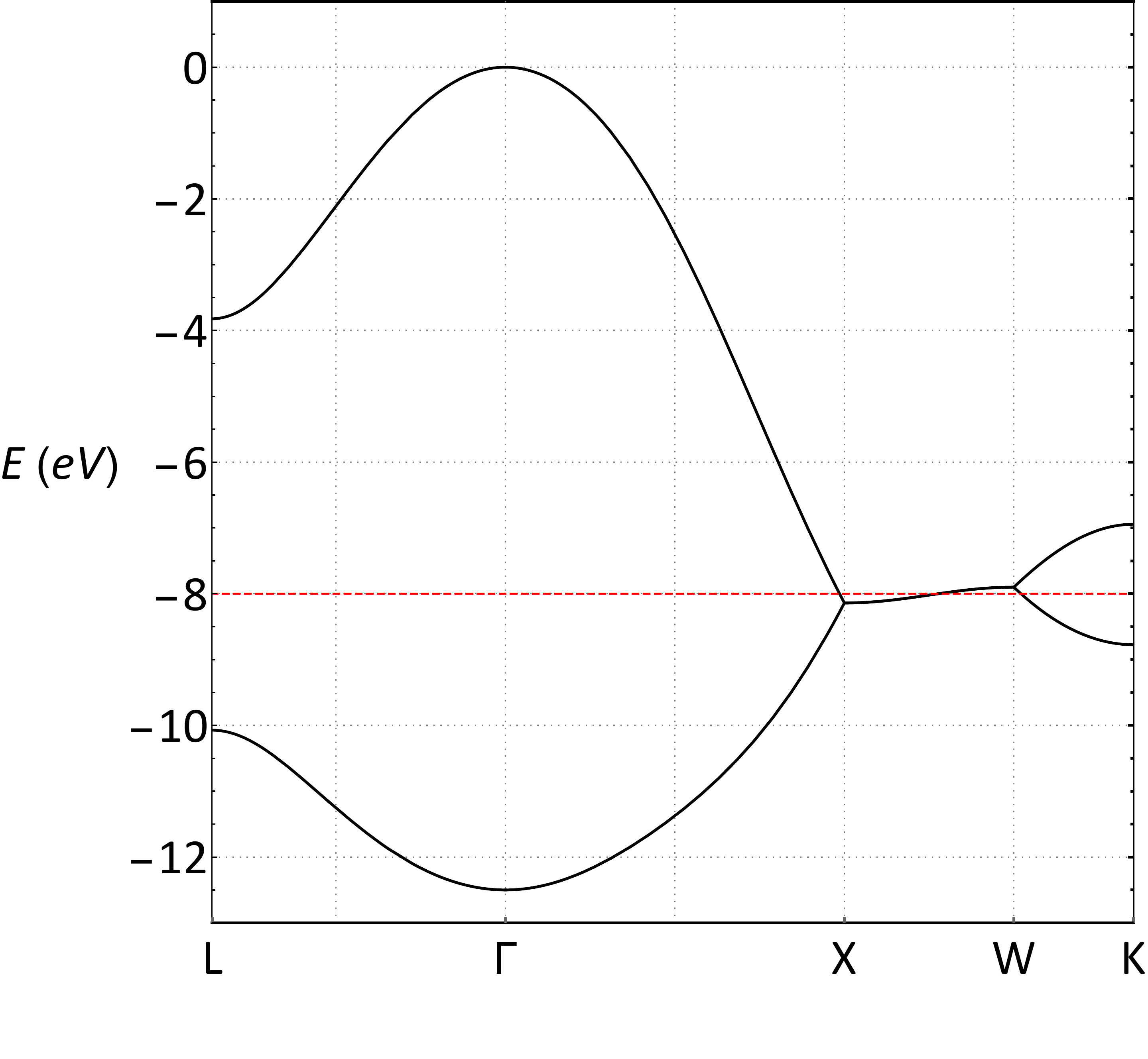}}
	\end{minipage}
	\hfill
	\begin{minipage}[h]{0.49\linewidth}
		\center{\includegraphics[width=1\linewidth]{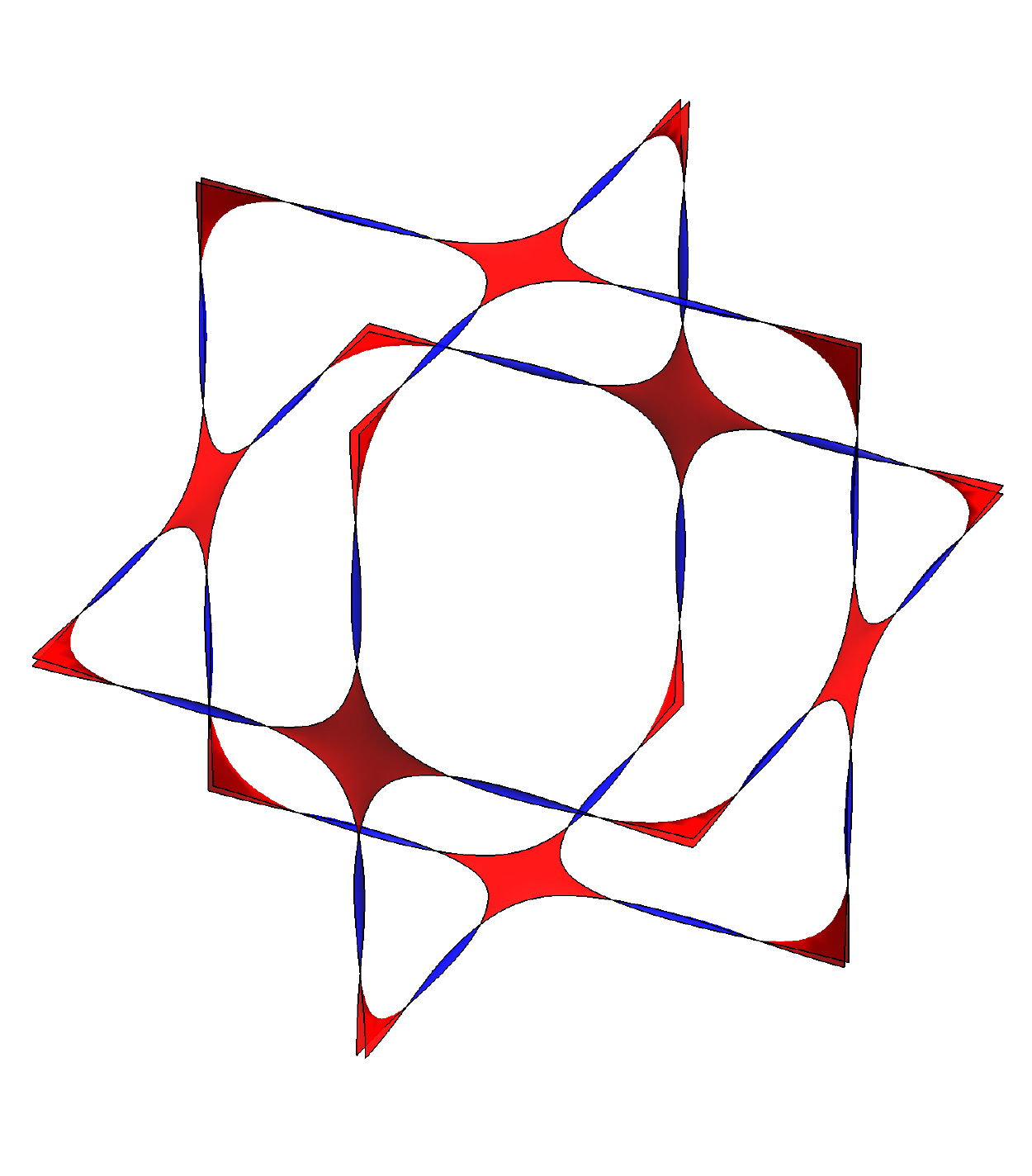}}
	\end{minipage}
	\caption{{\bf Electronic band structure and Fermi surface of the toy model.} \textit{Left:} Electronic band structure of toy model with next- and next-next-nearest neighbor hoppings included. The latter leads to nontrivial dispersion along X-W line. \textit{Right:} Fermi surface corresponding to the energy level shown with red dashed line on the left figure. The chosen energy crosses both bands, lower is shown with blue color and upper with red.}
	\label{fig:band_structure_toy}
\end{figure}

In the nearest neighbor toy model, chiral symmetry protects the
dispersionless nature of the wire frame and the Fermi Surface at this
energy coincides with the wire frame. The same happens at the touching
of the two lowest valence bands in the $sp^3s^*$ model, see
Fig.~\ref{fig:band_structure}. In real silicon, however, $X-W$ lines
possess nontrivial dispersion. Hopping matrix elements to farther
neighbors can be included in order to break sublattice symmetry and
reproduce this feature:
%
\begin{equation}
\label{eq:toy_2}
\begin{split}
H_{\text{toy-2}}(\kk)
=&
\hat{\sigma}_0\left(t_{n^3}v_{n^3}(\kk)+t_{n^4}v_{n^4}(\kk)\right)+
\\&+
t_{nn}\,
\left(\hat{\sigma}_1\;\Re g_0(\kk)+\hat{\sigma}_2\;\Im g_0(\kk)\right),
\end{split}
\end{equation}
where 
\begin{align}
	v_{n^3}(\kk)&=\frac{1}{2}\sum_{i\neq j}\cos\frac{k_ia}{2}\cos\frac{k_ja}{2},
	\\
	v_{n^4}(\kk)&=\sum_{i}\cos(k_ia).
\end{align}
We allow for both next-nearest-neighbor ($n^3$) and next-next-nearest-neighbor ($n^4$) hopping. It turns out that $n^3$ hopping alone is not sufficient to destroy the flatness of dispersion along X-W lines. We fit parameters of the toy model to reproduce $\Gamma, X, W$ point energies in the lowest valence band of real silicon. The resulting band structure and Fermi surface in the vicinity of the X point are shown in Fig.~\ref{fig:band_structure_toy}. 

\section{Silicon quantum well}

The clean and straightforward connection between bulk topology and
surface states relies on the sublattice symmetry, which,
unfortunately, is destroyed even within the $sp^3s^*$ model. While
$sp^3s^*$ model has matrix elements only between A and B sublattices,
only two of the ten states are involved in the formation of each wire
frame. When the full Hamiltonian is projected on the relevant two
dimensional subspace for each wire frame, longer range hopping
elements are induced via transitions to the states that were projected
out. For example, if we focus on the lowest energy wire frame that is
formed by the first two bands, next-nearest neighbor element between
an A site at the i-th and (i+1)-th cells is formed as
\begin{equation}
\label{eq:eff_H_perturbation}
\begin{split}
- \sum_{\alpha=2\dots10}\frac{\Braket{A_i,1|H_\text{eff}|B_{i},\alpha}\Braket{B_i,\alpha|H_\text{eff}|A_{i+1},1}}{(E_\alpha-E_1)^2}.
\end{split}
\end{equation}
%


\begin{figure*}[t!]
	\begin{minipage}[h]{0.405\linewidth}
		\center{\includegraphics[width=1\linewidth]{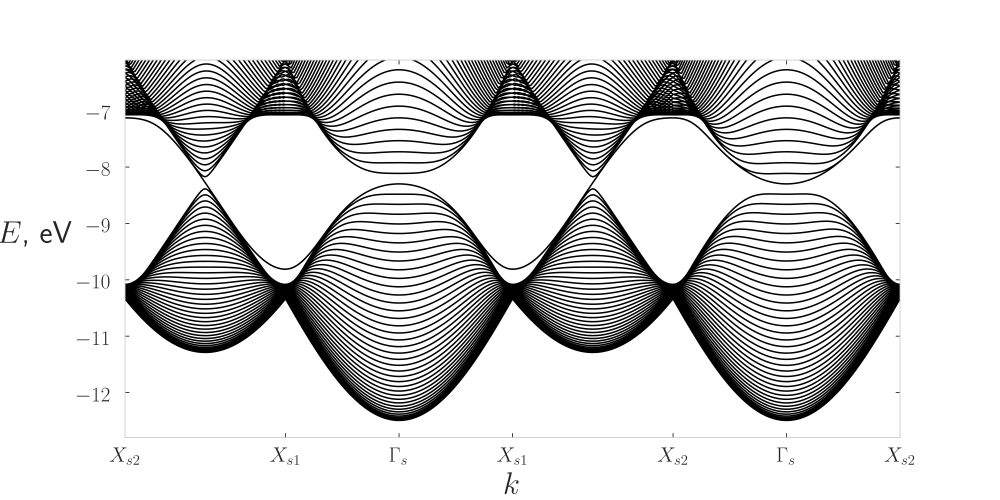}}
	\end{minipage}
	\hfill
	\begin{minipage}[h]{0.16\linewidth}
	\center{\includegraphics[width=1\linewidth]{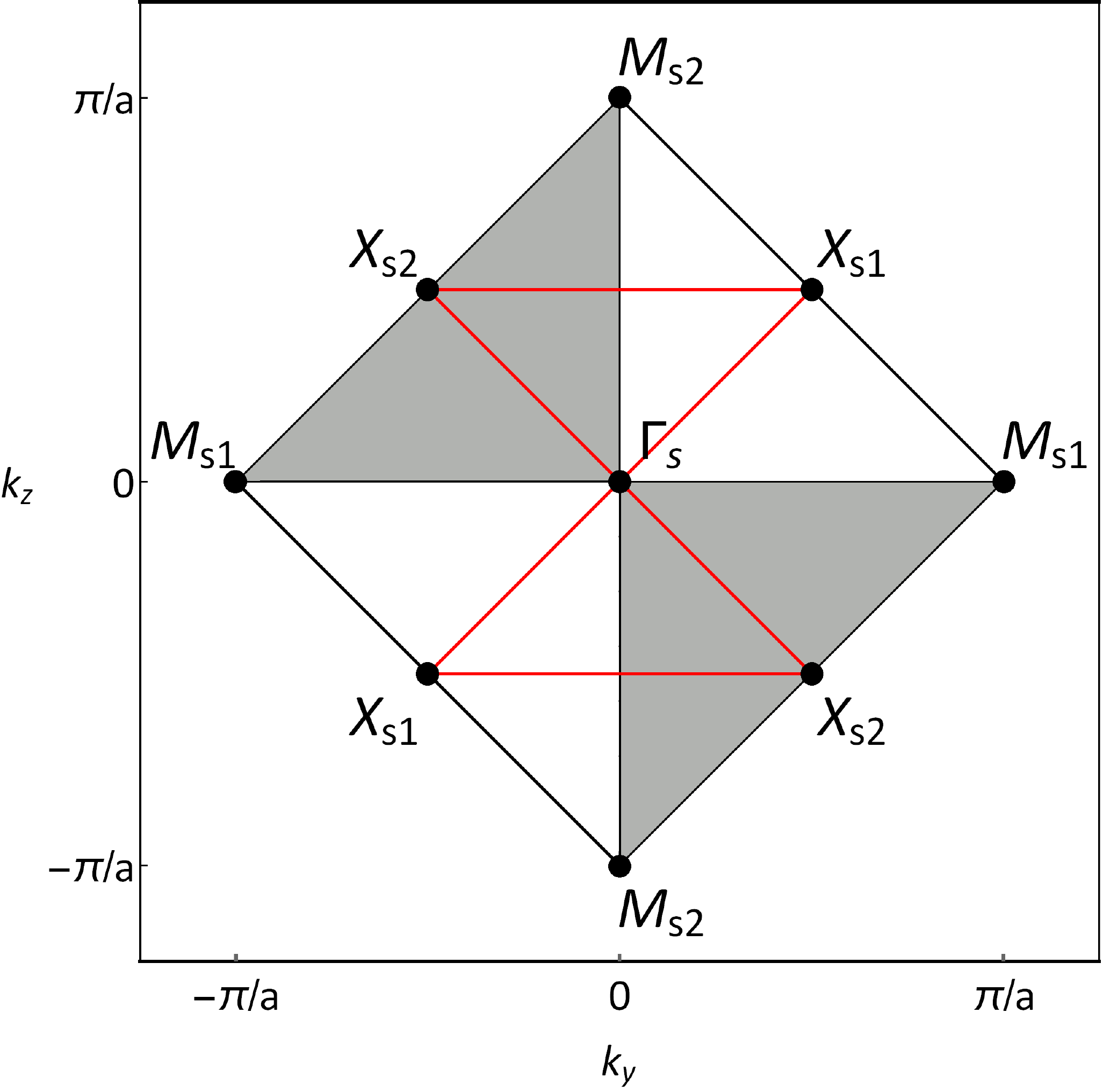}}
	\end{minipage}
	\hfill
	\begin{minipage}[h]{0.405\linewidth}
		\center{\includegraphics[width=1\linewidth]{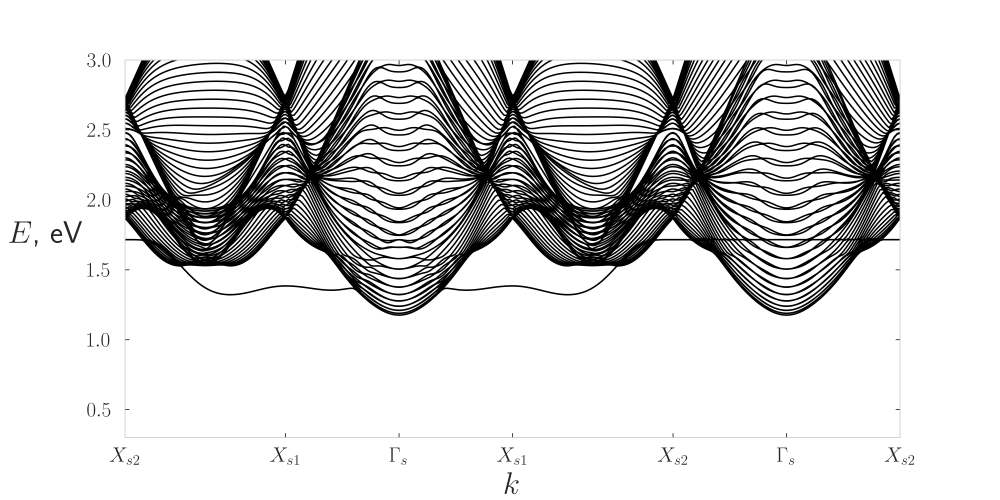}}
	\end{minipage}
	\caption{{\bf Energy dispersion $E(k_y,k_z)$} within a silicon slab (valence bands on the left and conduction bands on the  right). 
		Figure in the center shows surface Brillouin zone with shaded areas (2nd and 4th quadrants) revealing region where a surface state exists within a toy model.}
	\label{fig:slab_energy}
\end{figure*}

Despite this unfortunate fact, the bulk-boundary correspondence and surface states in the toy model are of \textit{topological} nature. Hence, we expect that breaking respective symmetries retains qualitatively bulk-boundary correspondence and surface states for reasonably large perturbations of the toy model.

To explore drumhead states within the $sp^3s^*$ model, we perform a numerical calculation of the electron dispersion in a silicon slab. We consider a slab of the  material with [100] surface and the thickness approximately 20 nm (40 unit cells). The sample is infinite in the $y,z$ directions parallel to the surface, for which we perform the Fourier transformation, retaining only the spatial dimension $x$ perpendicular to the surface, so that the electron wavefunction can be factorized as
\begin{equation}
	\Psi^{\alpha\mu}(x_i,k_y, k_z) = e^{i(k_yy+k_zz)}\psi^{\alpha\mu}_{k_y,k_z}(x_i), 
\end{equation}
where $x_i$ is the position of the $i$-th unit cell ($i =1\dots40$), $\alpha=1,2$ labels A/B sublattices and $\mu=1\dots5$ enumerates each of the five orbitals. We find all eigenvalues of the Hamiltonian $E_i(\kk_\parallel)$ for each value of the momentum $\kk_\parallel=(k_y,k_z)$ parallel to the surface and repeat this procedure along the path $\Gamma_\text{s}-\text{X}_\text{s1}-\text{X}_\text{s2}$ within the surface Brillouin zone. The result is shown on the Fig.~\ref{fig:slab_energy}. While the exact nature of the bulk-boundary correspondence does not survive in the five band model, the qualitative connection remains clear.

\begin{figure}[b!]
	\begin{minipage}[h]{.32\linewidth}
		\center{\includegraphics[width=1\linewidth]{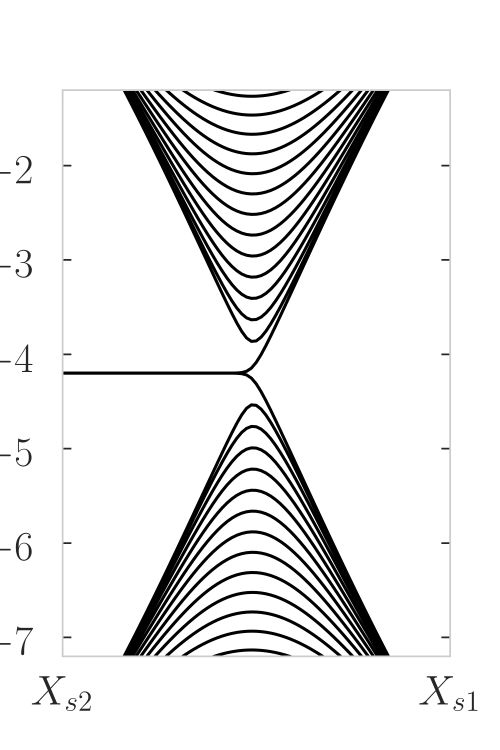}}
	\end{minipage}
	\hfill
	\begin{minipage}[h]{.32\linewidth}
		\center{\includegraphics[width=1\linewidth]{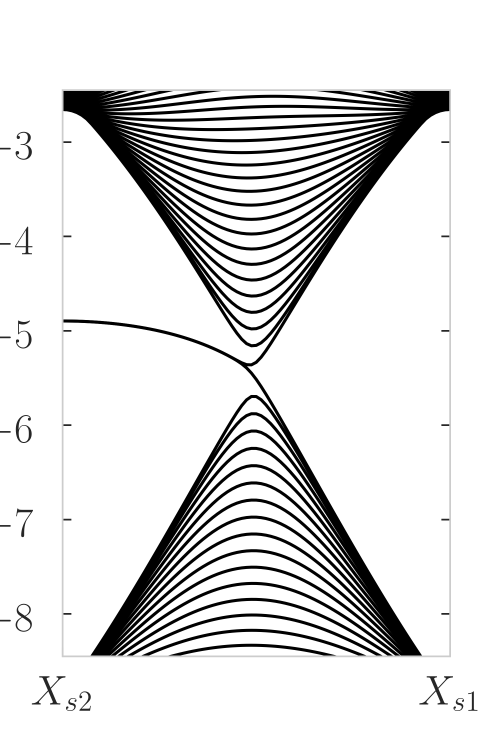}}
	\end{minipage}
	\hfill
	\begin{minipage}[h]{.32\linewidth}
		\center{\includegraphics[width=1\linewidth]{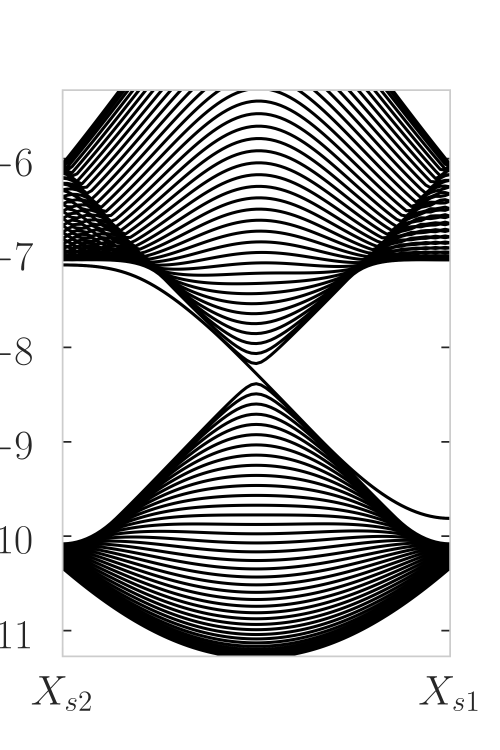}}
	\end{minipage}
	\begin{minipage}[h]{.32\linewidth}
		\center{\includegraphics[width=1\linewidth]{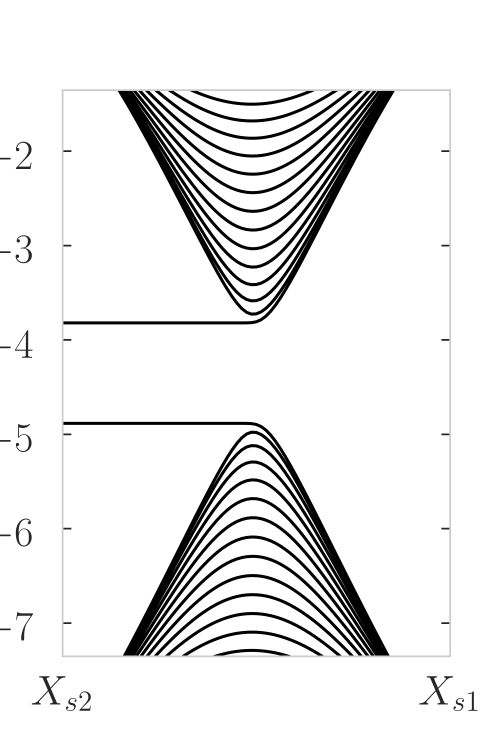}}
	\end{minipage}
	\hfill
	\begin{minipage}[h]{.32\linewidth}
		\center{\includegraphics[width=1\linewidth]{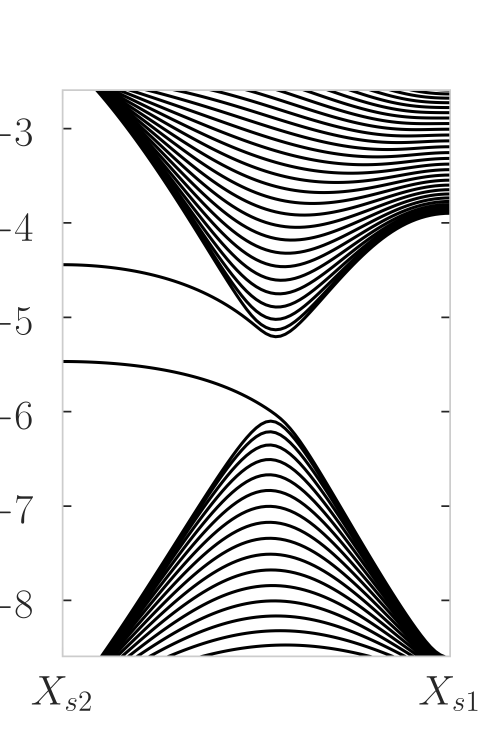}}
	\end{minipage}
	\hfill
	\begin{minipage}[h]{.32\linewidth}
		\center{\includegraphics[width=1\linewidth]{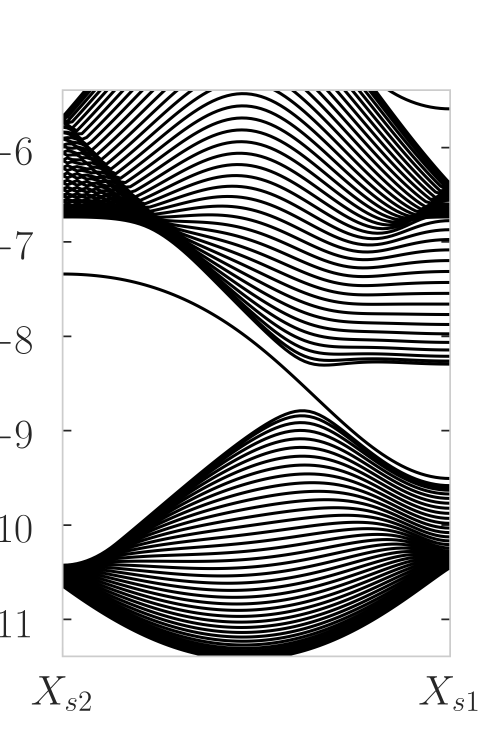}}
	\end{minipage}
	\caption{{\bf Drumhead state dispersion} for different interband interaction strengths $\alpha=0.0,\,0.5,\,1.0$ (columns). Bottom row shows dispersion with an additional A/B staggered chemical potential corresponding to opening of 1 eV bulk gap between the first and second valence bands. }
	\label{fig:band_crossing}
\end{figure}

As we have argued in the previous section, the energy dispersion of silicon in the vicinity of the touching of the lowest valence bands is fairly close to that of the toy model. Largely, this is the case due to other bands being well separated in energy from the first two valence bands. To further clarify the connection between the surface states and the bulk topology, we consider the following modification of the $sp^3s^*$ model. We slowly tune interband hopping parameters from 0$\%$ to $100\%$,
\begin{equation}
\label{eq:alpha}
\begin{pmatrix}
	V_{sp}^\prime(\alpha),&
V_{s^*p}^\prime(\alpha),& V_{xy}^\prime(\alpha)
\end{pmatrix}
=
\alpha\cdot
\begin{pmatrix}
V_{sp},&
V_{s^*p},& V_{xy}
\end{pmatrix},
\end{equation}
keeping other parameters intact. At $\alpha=0$ we have five exactly
solvable copies of the toy model and for each of them we know that the
drumhead states exist and they are of topological nature. This way, by
slowly tuning $\alpha$ from $0$ to $1$, we can track the evolution of
the surface state arising from the lowest energy wire frame.

In Fig.~\ref{fig:band_crossing} we zoom into the region between points
$X_{s2}$ and $X_{s1}$, to show a striking robustness of the drumhead
state in the full $sp^3s^*$ model. On the top panels, we show the
evolution of the spectrum for $\alpha=0, 0.5$, and 1.0, and on the
bottom panels we show the spectrum resulting from further addition of
a term that breaks inversion symmetry and opens a gap. Notice that the
drumhead states acquire a dispersion as $\alpha$ is turned on, and
that the corresponding bandwidth is large. Thus, if one opens a gap
that is smaller than that bandwidth (as in the bottom panels), a
stable drumhead state remains at the surface, crossing the bulk
gap. We remark that the scale of the bandwidth is rather large in the
$sp^3s^*$ model, of the order of several $eV$. In other words,
breaking of the chiral symmetry protects the drumhead state against
other perturbations that weakly break any other symmetry.


\begin{figure*}[t!]
	\begin{minipage}[h]{.185\linewidth}
		\center{\includegraphics[width=1\linewidth]{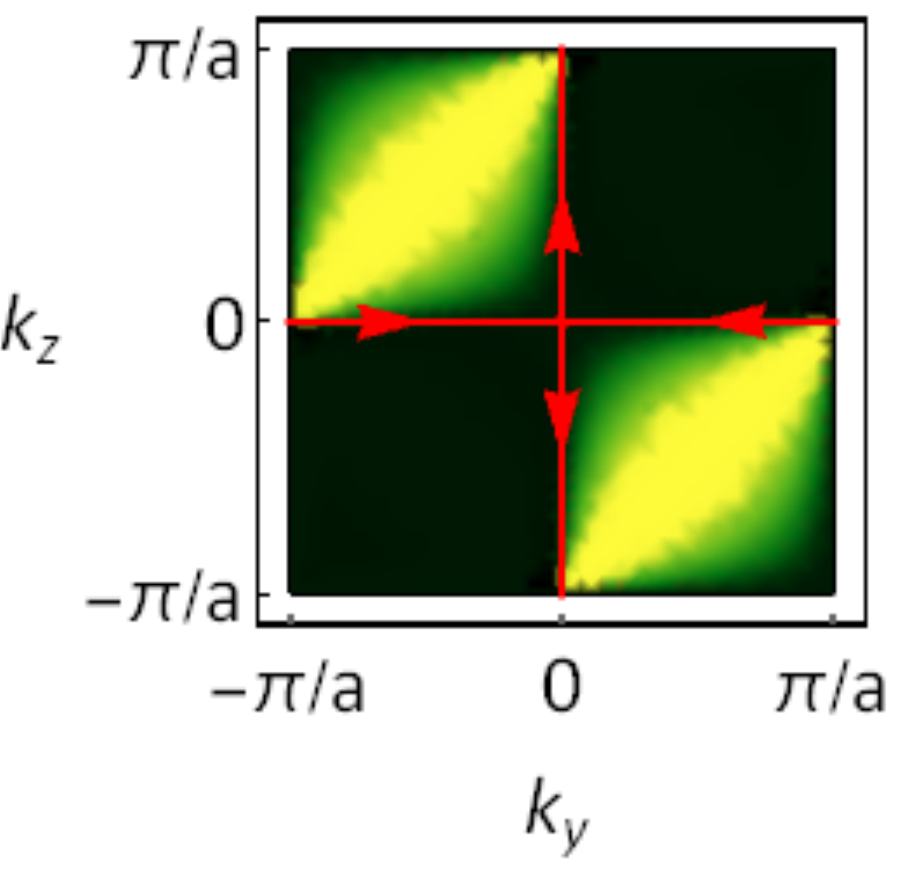}}
	\end{minipage}
	\hfill
	\begin{minipage}[h]{.185\linewidth}
		\center{\includegraphics[width=1\linewidth]{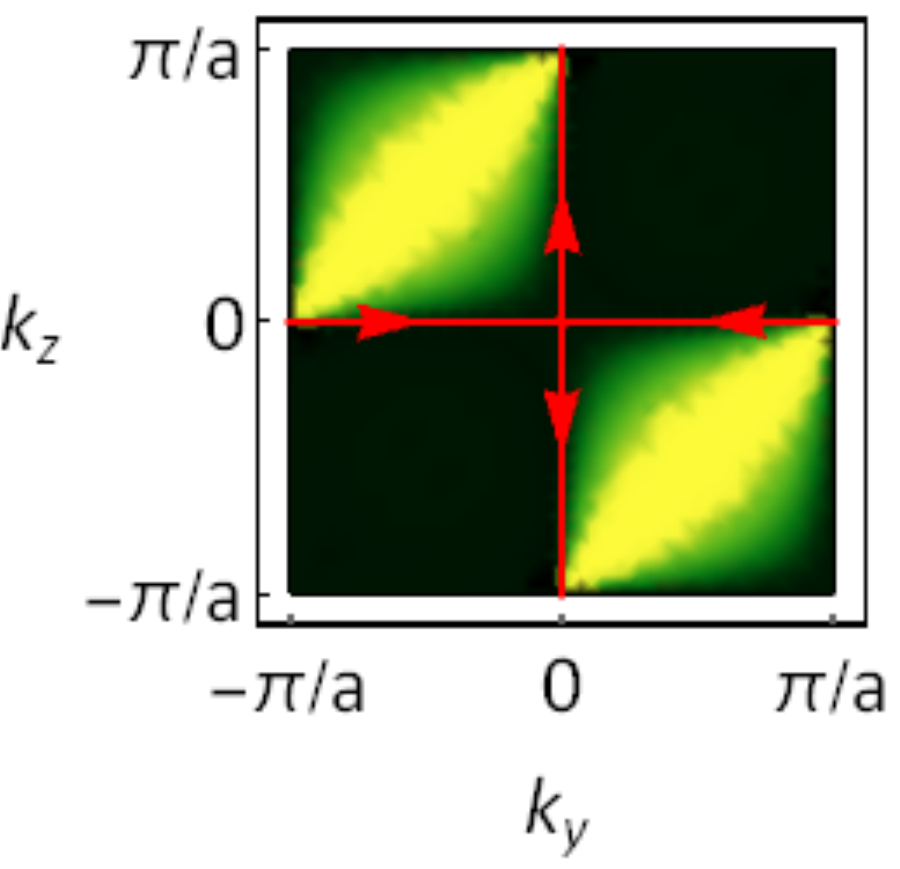}}
	\end{minipage}
	\hfill
	\begin{minipage}[h]{.185\linewidth}
		\center{\includegraphics[width=1\linewidth]{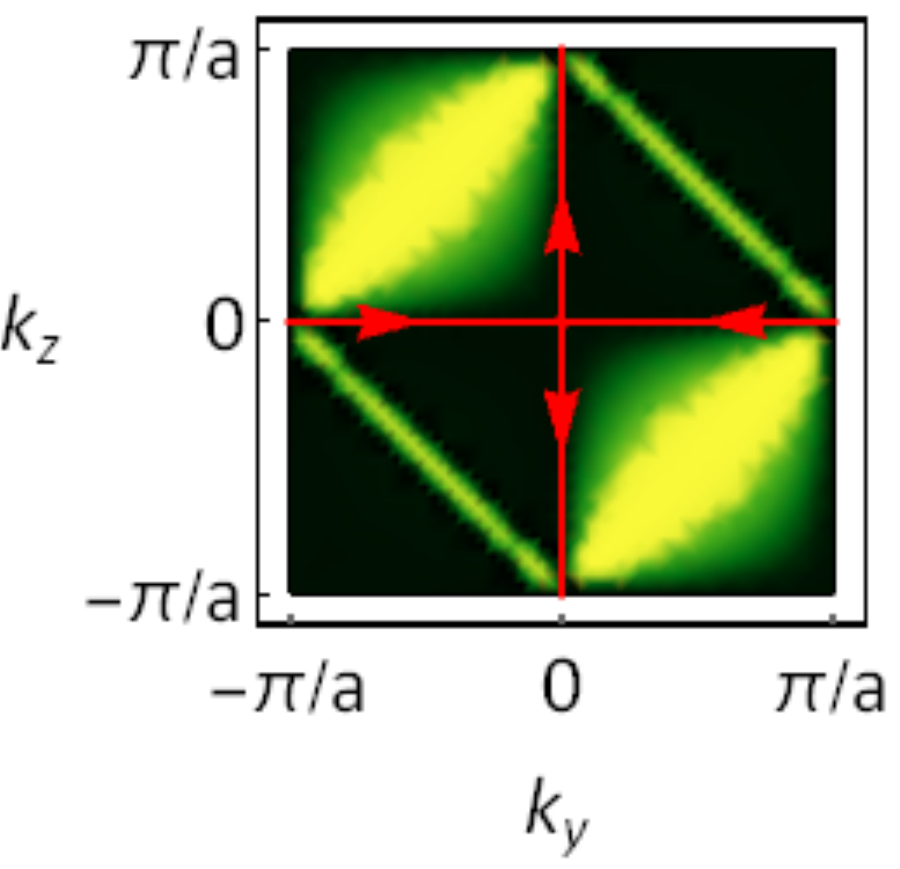}}
	\end{minipage}
	\hfill
	\begin{minipage}[h]{.185\linewidth}
		\center{\includegraphics[width=1\linewidth]{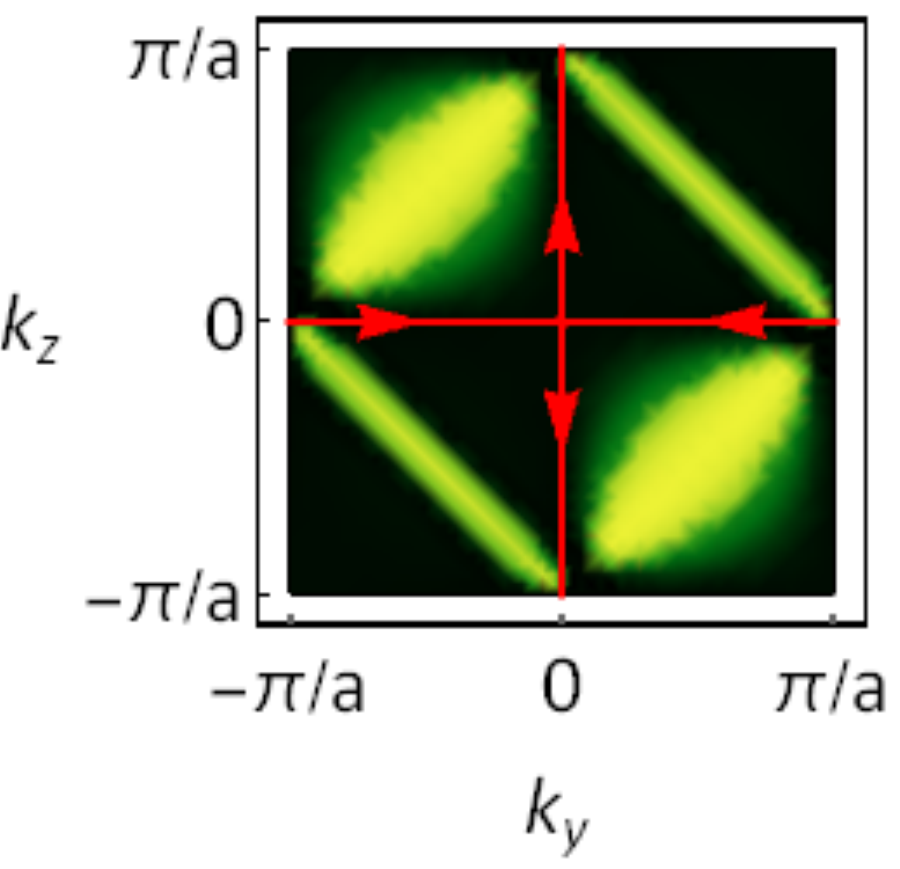}}
	\end{minipage}
	\hfill
	\begin{minipage}[h]{.185\linewidth}
		\center{\includegraphics[width=1\linewidth]{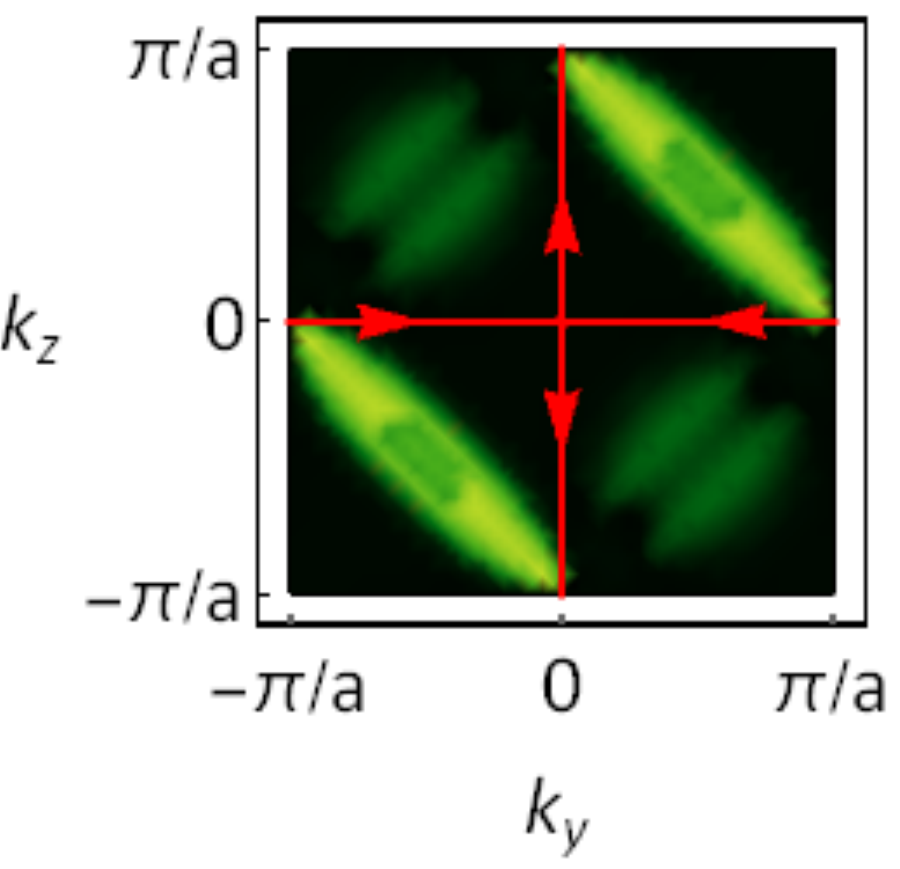}}
	\end{minipage}
	\hfill
	\begin{minipage}[h]{.04\linewidth}
		\center{\includegraphics[width=1\linewidth]{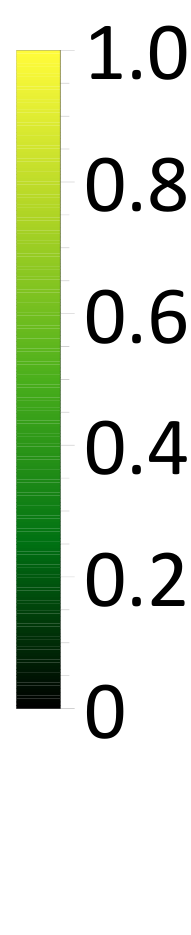}}
	\end{minipage}
	\caption{{\bf Inverse participation ratio} of the surface drumhead state related to the lowest in energy wire frame. The figures are given within the doubled surface Brillouin zone, and the progression from left to right represents tuning of interband interactions from complete absence $\alpha=0$ to a full $sp^3s^*$ model $\alpha=1$ (following Eq. (\ref{eq:alpha})). The figures are given for $\alpha=0,\,0.25,\,0.50,\,0.75,\,1$. The red cross shows a projection of the bulk wire frame onto the surface Brillouin zone.}
	\label{fig:ipr}
\end{figure*}

In addition to tracking the $\alpha$-dependence of the energy
spectrum, for each value of $\alpha$ we also calculate the inverse
participation ratio defined as
\begin{equation}
	\text{ipr}(k_y,k_z) = \sum_i \left|\psi_{k_y,k_z}(x_i)\right|^4.
\end{equation}
For localized states the ipr  is inversely proportional to the localization length $\propto1/l$, while for bulk states it vanishes as $\propto 1/L$, where $L$ is the size of the system.
The resulting inverse participation ratios are shown in the Fig.~\ref{fig:ipr}. Aside from the emergence of new surface states as we tune $\alpha$, the topological nature this surface state is unambiguous.



\section{Conclusions and Outlook}

In this work we identified a network of $\pi$ Berry flux lines in
reciprocal space, for silicon. We first constructed a rather general
argument, based solely on the existence of two sublattices, to argue
that there must be $\pi$ flux lines independently of how many bands
there are. The $\pi$ flux lines are tied to the spinor structure due
to the two sublattices, and the flux cannot spread out because of
time-reversal and inversion symmetry. The situation is analogous to
what happens in graphene, where the Dirac points carry $\pi$ flux and
cannot be removed perturbatively.

We then identified these lines in a tight-binding model containing
5-orbitals per sublattice (a $10 \times 10$ matrix Hamiltonian). We
showed explicitly that the $\pi$ flux lines appear, and identified the
$X$ point as a location where flux lines meet. The electronic
dispersion near the $X$ point can be described in terms of the Dirac
lines analyzed in this paper.

Flux lines inside the bulk Brillouin zone imply the existence of
drumhead surface states, which are confined within the projection of
the flux lines onto the surface Brillouin zone. We discussed how the
breaking of sublattice symmetry makes the identification of the
drumhead states less obvious, but that nevertheless the qualitative
connection between bulk-boundary remains.

There are two sets of questions that our work suggests: 
\begin{itemize}
	\item Can one observe all or some of these features experimentally?
	\item Can the knowledge that these topological features exist in silicon
	lead to novel electronics in this ``old'' material?
\end{itemize}

Regarding the first question, one of the possible tools to probe these
flux lines and the dispersion of wire frames is angle resolved
photoemission spectroscopy (ARPES). Since the features we identified
are present in all bands of silicon, one can study them in the valence
bands, which are accessible in ARPES. If high energy photons are used,
not only can one probe electronic states deep inside the valence bands
but also use the high penetration depth (here the $x$-direction) of
those photons to map constant energy surfaces as function of $k_y$ and
$k_z$ for different $k_x$. Such techniques are used in
Ref.~\onlinecite{Vladimir}, for example. These types of scans would be
able to identify the wire frames, providing evidence for the lines of
nodes where the flux runs through.

As for addressing the second question, one must find ways to pull the
physics of these nodal lines or surface states to the Fermi
level. While the Fermi level lies in the gap for intrinsic silicon,
one can reach the regime where the Fermi level crosses the conduction
band in inversion layers in doped silicon field effect transistors
(FETs).

Another possibility is to use undoped silicon, and pull down the
surface states by electric fields. States that are already localized
at the boundary are more sensitive to potentials caused by an electric
field (a linear potential has its largest and smallest values at the
boundary). Intrinsic silicon cannot screen the electric field. Once
the surface states are pulled to the Fermi level, there should be
metallic boundary states. The absence of disorder should lead to high
mobilities at these surfaces. While the mechanism described above does
not require that the surface states be drumhead ones, we already
observe from Fig.~\ref{fig:slab_energy} that the lowest surface band
at positive energies is of the drumhead type.

The findings presented in this work reveal novel topological
electronic features in the band structure of silicon, one of the best
known and most studied materials. That these features had been missed
does not signal an accident, but rather suggests that there are a
number of topological properties occuring in many, if not most, other
materials. The topological features of silicon that we expose
provide new impetus to revisit the physics of bulk silicon and
two-dimensional electron gases in silicon FETs, particularly in light
of what is now known from recent studies of both graphene and
topological insulators.




\begin{acknowledgements}
  We thank Michael El-Batanouny, Shyam Erramilli, Thomas Iadecola,
  Christopher Mudry, Andrei Ruckenstein, and Alex Sushkov for
  insightful discussions. We acknowledge financial support from the
  U.S. Dept. of Energy Grant DE-FG02-06ER46316 (C.C.).
\end{acknowledgements}



\bibliography{silicon}

\begin{thebibliography}{10}

\bibitem{vonKlitzing80}
K.~v. Klitzing, G.~Dorda, and M.~Pepper.
\newblock New method for high-accuracy determination of the fine-structure
  constant based on quantized {Hall} resistance.
\newblock {\em Phys. Rev. Lett.}, 45:494--497, Aug 1980.

\bibitem{TKNN82}
D.~J. Thouless, M.~Kohmoto, M.~P. Nightingale, and M.~den Nijs.
\newblock Quantized {Hall} conductance in a two-dimensional periodic potential.
\newblock {\em Phys. Rev. Lett.}, 49:405--408, Aug 1982.

\bibitem{Kane05a}
C.~L. Kane and E.~J. Mele.
\newblock Quantum spin {Hall} effect in graphene.
\newblock {\em Phys. Rev. Lett.}, 95:226801, Nov 2005.

\bibitem{Kane05b}
C.~L. Kane and E.~J. Mele.
\newblock $\mathbb{Z}_{2}$ topological order and the quantum spin {Hall}
  effect.
\newblock {\em Phys. Rev. Lett.}, 95:146802, Sep 2005.

\bibitem{Fu-Kane06}
Liang Fu and C.~L. Kane.
\newblock Time reversal polarization and a $\mathbb{Z}_{2}$ adiabatic spin
  pump.
\newblock {\em Phys. Rev. B}, 74:195312, Nov 2006.

\bibitem{Bernevig06a}
B.~Andrei Bernevig and Shou-Cheng Zhang.
\newblock Quantum spin {Hall} effect.
\newblock {\em Phys. Rev. Lett.}, 96:106802, Mar 2006.

\bibitem{Bernevig06b}
B.~Andrei Bernevig, Taylor~L. Hughes, and Shou-Cheng Zhang.
\newblock Quantum spin {Hall} effect and topological phase transition in
  {H}g{T}e quantum wells.
\newblock {\em Science}, 314(5806):1757--1761, 2006.

\bibitem{Konig07}
Markus K{\"o}nig, Steffen Wiedmann, Christoph Br{\"u}ne, Andreas Roth, Hartmut
  Buhmann, Laurens~W. Molenkamp, Xiao-Liang Qi, and Shou-Cheng Zhang.
\newblock Quantum spin {Hall} insulator state in {HgTe} quantum wells.
\newblock {\em Science}, 318(5851):766--770, 2007.

\bibitem{Fu07}
Liang Fu, C.~L. Kane, and E.~J. Mele.
\newblock Topological insulators in three dimensions.
\newblock {\em Phys. Rev. Lett.}, 98:106803, Mar 2007.

\bibitem{Moore07}
J.~E. Moore and L.~Balents.
\newblock Topological invariants of time-reversal-invariant band structures.
\newblock {\em Phys. Rev. B}, 75:121306, Mar 2007.

\bibitem{Roy09a}
Rahul Roy.
\newblock $\mathbb{Z}_{2}$ classification of quantum spin {Hall} systems: An
  approach using time-reversal invariance.
\newblock {\em Phys. Rev. B}, 79:195321, May 2009.

\bibitem{Roy09b}
Rahul Roy.
\newblock Topological phases and the quantum spin {Hall} effect in three
  dimensions.
\newblock {\em Phys. Rev. B}, 79:195322, May 2009.

\bibitem{Hsieh08}
D.~Hsieh, D.~Qian, L.~Wray, Y.~Xia, Y.~S. Hor, R.~J. Cava, and M.~Z. Hasan.
\newblock A topological {Dirac} insulator in a quantum spin {Hall} phase.
\newblock {\em Nature}, 452(7190):970--974, Apr 2008.

\bibitem{Hsieh09}
D.~Hsieh, Y.~Xia, L.~Wray, D.~Qian, A.~Pal, J.~H. Dil, J.~Osterwalder,
  F.~Meier, G.~Bihlmayer, C.~L. Kane, Y.~S. Hor, R.~J. Cava, and M.~Z. Hasan.
\newblock Observation of unconventional quantum spin textures in topological
  insulators.
\newblock {\em Science}, 323(5916):919--922, 2009.

\bibitem{Hsieh09b}
D.~Hsieh, Y.~Xia, D.~Qian, L.~Wray, J.~H. Dil, F.~Meier, J.~Osterwalder,
  L.~Patthey, J.~G. Checkelsky, N.~P. Ong, A.~V. Fedorov, H.~Lin, D.~Bansil, A.
  eand~Grauer, Y.~S. Hor, R.~J. Cava, and M.~Z. Hasan.
\newblock A tunable topological insulator in the spin helical {Dirac} transport
  regime.
\newblock {\em Nature}, 460(7259):1101--1105, Aug 2009.

\bibitem{Chen09}
Y.~L. Chen, J.~G. Analytis, J.-H. Chu, Z.~K. Liu, S.-K. Mo, X.~L. Qi, H.~J.
  Zhang, D.~H. Lu, X.~Dai, Z.~Fang, S.~C. Zhang, I.~R. Fisher, Z.~Hussain, and
  Z.-X. Shen.
\newblock Experimental realization of a three-dimensional topological
  insulator, {B}i$_{2}${T}e$_{3}$.
\newblock {\em Science}, 325(5937):178--181, 2009.

\bibitem{Hasan-Kane10}
M.~Z. Hasan and C.~L. Kane.
\newblock Colloquium: Topological insulators.
\newblock {\em Rev. Mod. Phys.}, 82:3045--3067, Nov 2010.

\bibitem{Qi-Zhang11}
Xiao-Liang Qi and Shou-Cheng Zhang.
\newblock Topological insulators and superconductors.
\newblock {\em Rev. Mod. Phys.}, 83:1057--1110, Oct 2011.

\bibitem{BurkovHookBalents}
A.~A. Burkov, M.~D. Hook, and Leon Balents.
\newblock Topological nodal semimetals.
\newblock {\em Phys. Rev. B}, 84:235126, Dec 2011.

\bibitem{BurkovBalents}
A.~A. Burkov and Leon Balents.
\newblock Weyl semimetal in a topological insulator multilayer.
\newblock {\em Phys. Rev. Lett.}, 107:127205, Sep 2011.

\bibitem{Wan11}
Xiangang Wan, Ari~M. Turner, Ashvin Vishwanath, and Sergey~Y. Savrasov.
\newblock Topological semimetal and {Fermi}-arc surface states in the
  electronic structure of pyrochlore iridates.
\newblock {\em Phys. Rev. B}, 83:205101, May 2011.

\bibitem{Neupane2014}
Madhab Neupane, Su-Yang Xu, Raman Sankar, Nasser Alidoust, Guang Bian, Chang
  Liu, Ilya Belopolski, Tay-Rong Chang, Horng-Tay Jeng, Hsin Lin, Arun Bansil,
  Fangcheng Chou, and M.~Zahid Hasan.
\newblock Observation of a three-dimensional topological {Dirac} semimetal
  phase in high-mobility {C}d$_3${A}s$_2$.
\newblock {\em Nature Communications}, 5:3786 EP --, May 2014.
\newblock Article.

\bibitem{Liu14}
Z.~K. Liu, B.~Zhou, Y.~Zhang, Z.~J. Wang, H.~M. Weng, D.~Prabhakaran, S.-K. Mo,
  Z.~X. Shen, Z.~Fang, X.~Dai, Z.~Hussain, and Y.~L. Chen.
\newblock Discovery of a three-dimensional topological {Dirac} semimetal,
  {N}a$_3${B}i.
\newblock {\em Science}, 343(6173):864--867, 2014.

\bibitem{Huang2015}
Shin-Ming Huang, Su-Yang Xu, Ilya Belopolski, Chi-Cheng Lee, Guoqing Chang,
  BaoKai Wang, Nasser Alidoust, Guang Bian, Madhab Neupane, Chenglong Zhang,
  Shuang Jia, Arun Bansil, Hsin Lin, and M.~Zahid Hasan.
\newblock A {W}eyl {Fermion} semimetal with surface {Fermi} arcs in the
  transition metal monopnictide {T}a{A}s class.
\newblock {\em Nature Communications}, 6:7373 EP --, Jun 2015.
\newblock Article.

\bibitem{Lv15}
B.~Q. Lv, H.~M. Weng, B.~B. Fu, X.~P. Wang, H.~Miao, J.~Ma, P.~Richard, X.~C.
  Huang, L.~X. Zhao, G.~F. Chen, Z.~Fang, X.~Dai, T.~Qian, and H.~Ding.
\newblock Experimental discovery of {W}eyl semimetal {T}a{A}s.
\newblock {\em Phys. Rev. X}, 5:031013, Jul 2015.

\bibitem{Yu15}
Rui Yu, Hongming Weng, Zhong Fang, Xi~Dai, and Xiao Hu.
\newblock Topological node-line semimetal and {Dirac} semimetal state in
  antiperovskite {C}u$_{3}${P}d{N}.
\newblock {\em Phys. Rev. Lett.}, 115:036807, Jul 2015.

\bibitem{bzdusek_etal_2016}
Tom\'a\v{s} Bzdu\v{s}ek, QuanSheng Wu, Andreas Ruegg, Manfred Sigrist, and
  Alexey~A. Soluyanov.
\newblock Nodal-chain metals.
\newblock {\em Nature}, 538(7623):75--78, OCT 6 2016.

\bibitem{Schnyder08}
Andreas~P. Schnyder, Shinsei Ryu, Akira Furusaki, and Andreas W.~W. Ludwig.
\newblock Classification of topological insulators and superconductors in three
  spatial dimensions.
\newblock {\em Phys. Rev. B}, 78:195125, Nov 2008.

\bibitem{VOGL1983}
P.~Vogl, Harold~P. Hjalmarson, and John~D. Dow.
\newblock A semi-empirical tight-binding theory of the electronic structure of
  semiconductors.
\newblock {\em Journal of Physics and Chemistry of Solids}, 44(5):365 -- 378,
  1983.

\bibitem{Berry_gauge_choice}
{The Berry phase can be read directly from the product of unitary matrices
  because of our choice of gauge. Having the phases $\phi(\GG_i)= 2 \pi /N$
  (for $N=3$ in graphene and $N=4$ in silicon) ensures that $\phi(\GG_i)/2
  =\pi/N <\pi/2$, and the overlap between spinors along the legs of the walk is
  real and positive, {\it i.e.}, we are working with the gauge of parallel
  transport. In this case, the Berry phase can be read from the total rotation
  of the spinor from beginning to end.}

\bibitem{Slater_book}
J.C. Slater.
\newblock {\em Symmetry and Energy Bands in Crystals}.
\newblock Dover Publications, 1965.

\bibitem{ioffe}
New {S}emiconductor {M}aterials archive: {P}hysical {P}roperties of
  {S}emiconductors ({Si}).
\newblock \url{http://www.ioffe.ru/SVA/NSM/Semicond/Si/bandstr.html}.

\bibitem{yan2018experimental}
Qinghui Yan, Rongjuan Liu, Zhongbo Yan, Boyuan Liu, Hongsheng Chen, Zhong Wang,
  and Ling Lu.
\newblock Experimental discovery of nodal chains.
\newblock {\em Nature Physics}, page~1, 2018.

\bibitem{atala2013}
Marcos Atala, Monika Aidelsburger, Julio~T Barreiro, Dmitry Abanin, Takuya
  Kitagawa, Eugene Demler, and Immanuel Bloch.
\newblock Direct measurement of the zak phase in topological bloch bands.
\newblock {\em Nature Physics}, 9(12):795, 2013.

\bibitem{rhim2017bulk}
Jun-Won Rhim, Jan Behrends, and Jens~H Bardarson.
\newblock Bulk-boundary correspondence from the intercellular zak phase.
\newblock {\em Physical Review B}, 95(3):035421, 2017.

\bibitem{van2016topological}
Guido van Miert, Carmine Ortix, and Cristiane~Morais Smith.
\newblock Topological origin of edge states in two-dimensional
  inversion-symmetric insulators and semimetals.
\newblock {\em 2D Materials}, 4(1):015023, 2016.

\bibitem{naumov2016topological}
Ivan~I Naumov and Russell~J Hemley.
\newblock Topological surface states in dense solid hydrogen.
\newblock {\em Physical Review Letters}, 117(20):206403, 2016.

\bibitem{Vladimir}
Vladimir~N. Strocov, Ming Shi, Masaki Kobayashi, Claude Monney, Xiaoqiang Wang,
  Juraj Krempasky, Thorsten Schmitt, Luc Patthey, Helmuth Berger, and Peter
  Blaha.
\newblock {Three-Dimensional Electron Realm in ${\mathrm{VSe}}_{2}$ by
  Soft-X-Ray Photoelectron Spectroscopy: Origin of Charge-Density Waves}.
\newblock {\em Phys. Rev. Lett.}, 109:086401, Aug 2012.

\end{thebibliography}





\newpage
\begin{widetext}

\appendix
\setcounter{equation}{0}
\renewcommand\theequation{S\arabic{equation}}
\begin{center}
	Supplementary Online Information
\end{center}

\section{Spinor structure and Berry fluxes in the graphene and silicon lattices}

Let us first apply the formalism presented in the main text to derive
the familiar result that in graphene there is a $\pi$-vortex in the
Brillouin zone at the K point. The carbon atoms form a honeycomb
lattice, with interpenetrating triangular sublattices $A$ and $B$. The
$A$ sites sit on lattice sites spanned by the basis vectors
$\mathbf{R}_1=\sqrt{3}\,a\, (1/2, \sqrt{3}/2)$ and 
$\mathbf{R}_2=\sqrt{3}\,a\, (-1/2, \sqrt{3}/2)$, where $a=1.42$\AA.
The reciprocal lattice vectors are
$\GG_1=\frac{4\pi}{3a}\, (\sqrt{3}/2, 1/2)$ and 
$\GG_2=\frac{4\pi}{3a}\, (-\sqrt{3}/2, 1/2)$.
The three vectors connecting sublattice $A$ to $B$ are
$\mathbf{d}_0=\, a\,(0, 1)$,
$\mathbf{d}_1=\, a\,(-\sqrt{3}/2, -1/2)$, and
$\mathbf{d}_2=\, a\,(\sqrt{3}/2, -1/2)$.

The off-diagonal hopping matrix element for the $\pi$-orbital is
\begin{equation}
  H^{\;}_{AB}(\kk)
  =
  -
  \sum_{\mu=0}^2\;
  t(\mathbf{d}_\mu)\;e^{\im\kk\cdot\mathbf{d}_\mu}
     \;,
\end{equation}
where we allowed generically for unequal values for the hoppings to
the three nearest neighbors. The vectors $\GG_{1,2}$ and
$\mathbf{d}_{0,1,2}$ satisfy
\begin{equation}
  \GG_{i} \cdot \mathbf{d}_\mu = \frac{2\pi}{3}\;\;(\text{mod}
  \;2\pi)\quad \text{for}\; i=1,2\; \text{and}\; \mu=0,1,2 \;.
\end{equation}
It then follows that the $2\times 2$ Hamiltonian for graphene is not
periodic in $\kk$, but instead is periodic up to the unitary
transformation in Eq.~(\ref{eq:H-gauge}) of the main text (repeated
here for convenience):
\begin{equation}
  H(\kk+\GG_i)
  =
  U(\GG_i)\;H(\kk)\;U^\dagger(\GG_i)
  \;,
  \quad
  \text{with}\quad
  U(\GG_i)
  =
  e^{\;\im\frac{1}{2}\Phi(\GG_i)\;\sigma_z}
  \;,
  \label{eq:H-gauge-repeated}
\end{equation}
with $\Phi(\GG_i)=2\pi/3$. One can also check that
$-(\GG_{1}+\GG_{2}) \cdot \mathbf{d}_\mu =
\frac{2\pi}{3}\;(\text{mod} \;2\pi)$, for $\mu=0,1,2$, or
equivalently, $\Phi(-\GG_1-\GG_2)=2\pi/3$.

We then consider the walk in $\kk$-space that visits, in order, the
points $\kk,\kk+\GG_1,\kk+\GG_1+\GG_2$ and back to $\kk$. This walk visits
three neighboring Brillouin zones, returning to the original point in
$\kk$-space, closing a loop, shown in Fig.~\ref{fig:recip_lattice}.
We know that, in graphene, this loop will not visit a degeneracy point
if we choose $\kk=0$ (the $\Gamma$ point). (In silicon, we shift the
point $\kk$ from the origin to avoid going through degeneracies.) At
the end of the walk, the Hamiltonian returns to $H(\kk)$, but the
eigenvector is rotated by the sequence of unitary operations
\begin{equation}
  \begin{split}
    \label{eq:3_Us}
  U(-\GG_1-\GG_2)\;U(\GG_2)\;U(\GG_1)
  &=
  e^{\;\im\frac{1}{2}\left[
    \Phi(-\GG_1-\GG_2)+\Phi(\GG_2)+\Phi(\GG_1)
    \right]\;\sigma_z}
  \\
  &=
  e^{\;\im\pi\;\sigma_z}
  =
  -\oneone
  \;.
  \end{split}
\end{equation}
This rotation changes the sign of the wavefunction, which is
equivalent to an accumulation of a Berry phase of
$\pi$~\,\cite{Berry_gauge_choice}. Indeed, the Dirac node at the K
point contains the $\pi$-vortex when all the three hoppings to the
nearest neighbors are equal. If these three hoppings are not equal,
the vortex moves location in $\kk$-space, but cannot disappear; it
must be contained within the triangle. To disappear, the vortex must
come to the boundaries of the triangle to meet an anti-vortex, but
when that happens our assumption that the path does not include a
degenerate point no longer applies.

\begin{figure}
	\begin{minipage}[h]{.49\linewidth}
		\center{\includegraphics[width=0.8\linewidth]{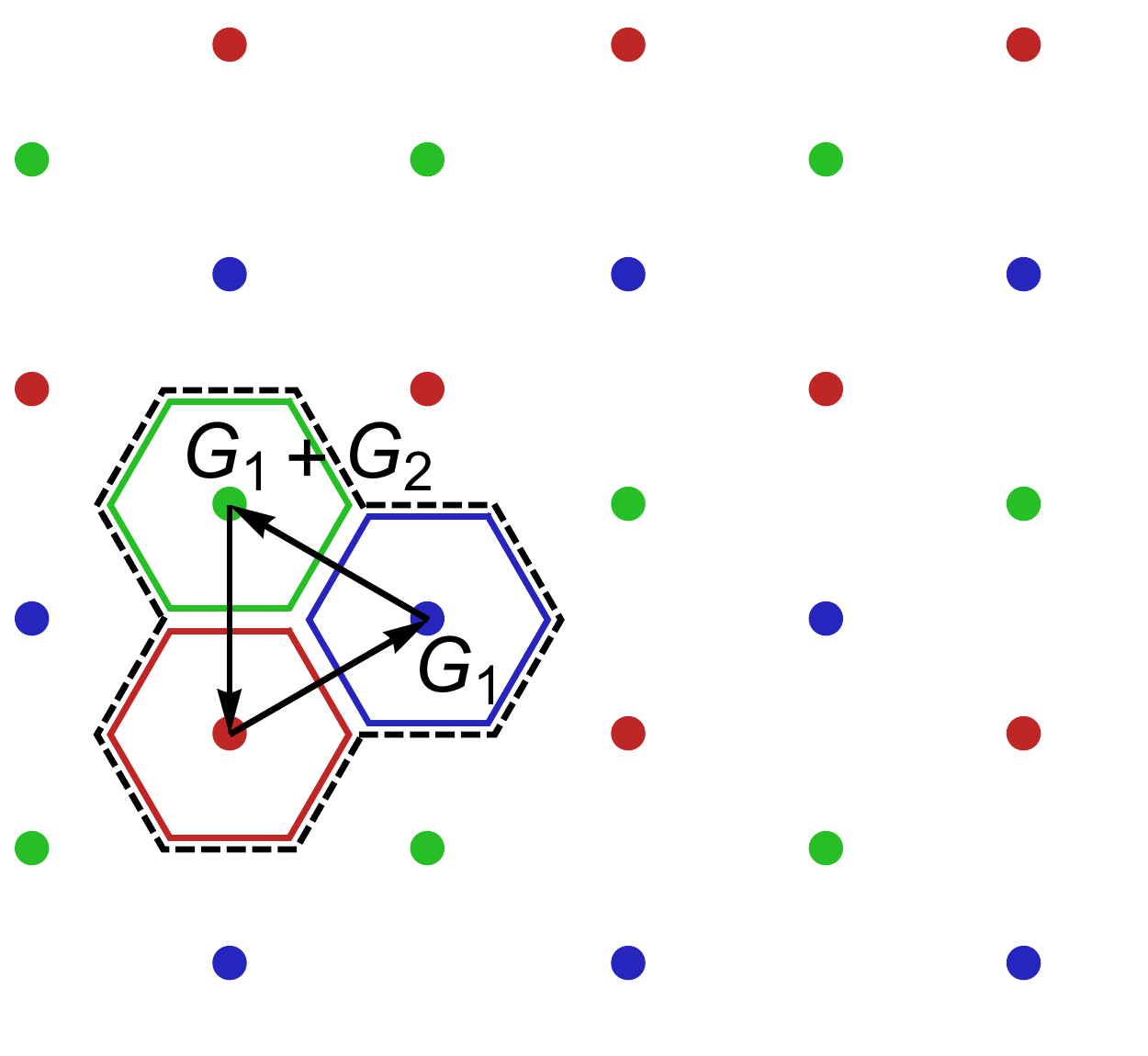}}
	\end{minipage}
	\hfill
	\begin{minipage}[h]{.49\linewidth}
		\center{\includegraphics[width=0.8\linewidth]{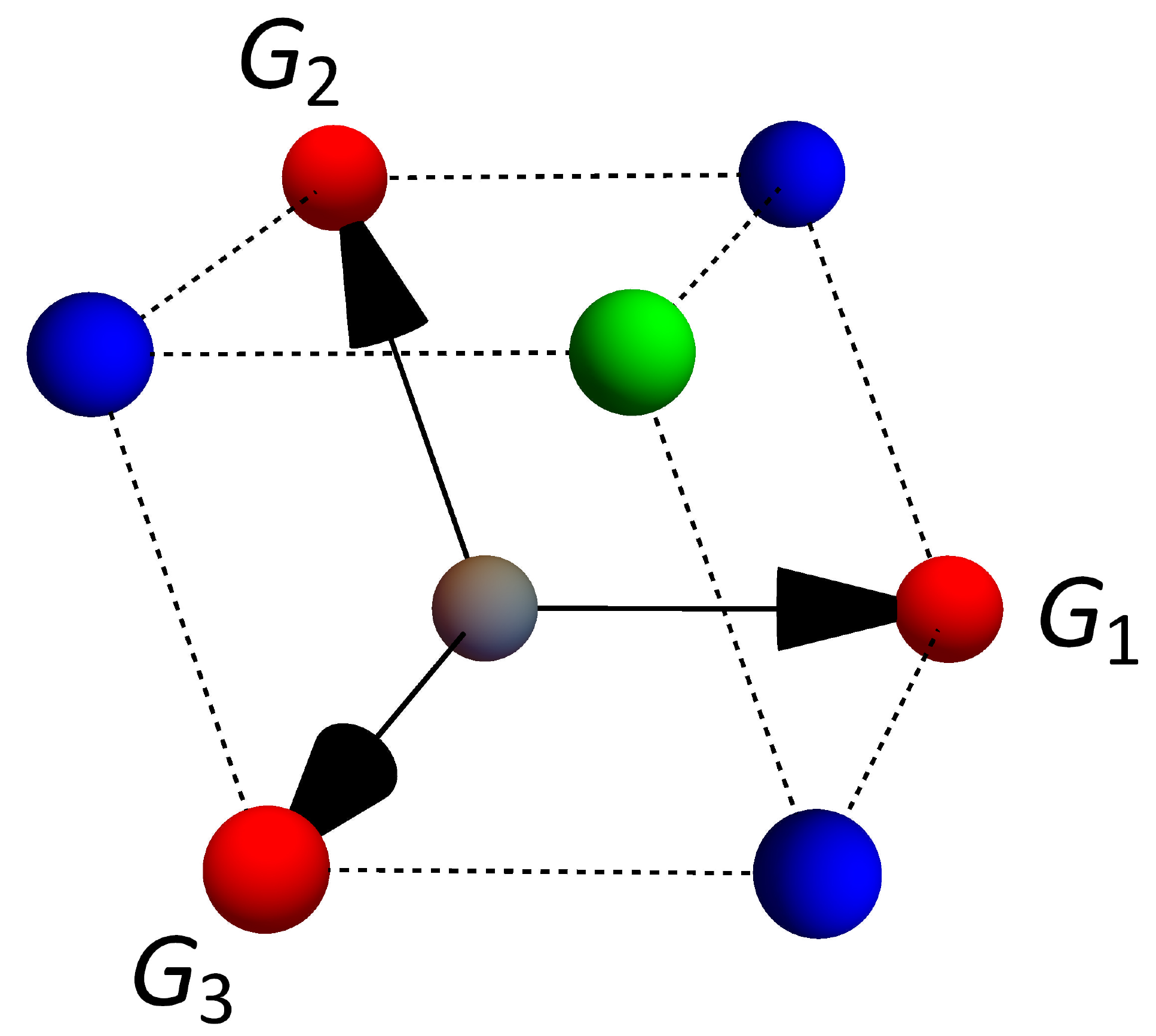}}
	\end{minipage}
	\caption{{\bf Reciprocal lattice.} \textit{Left:} reciprocal lattice of graphene. There are three possible axial phases at the center of the Brillouin zone, $0, 2\pi/3, 4\pi/3$, which are represented by three colors. This leads to tripling of the Brillouin zone, as shown by the dashed line emclosing three hexagons of different colors. The figure also shows a loop constructed out of $\GG$ vectors with an overall $\pi$ Berry phase that establishes existence of the Dirac points at the K/K' points. 
	\newline\textit{Right:} reciprocal lattice of silicon. In the case of silicon there are four possible axial phases and hence four colors for the reciprocal lattice sites, indicating the quadrupling of the Brillouin zone.}
	\label{fig:recip_lattice}
\end{figure}

Having completed the warm up exercise of recovering know results for
graphene in two spatial dimensions, we apply the same steps for
silicon in three dimensions. In silicon the $A$ sites sit on lattice
sites spanned by the basis vectors
$\mathbf{R}_1=\frac{a}{2}\,(0,1,1)$,
$\mathbf{R}_2=\frac{a}{2}\,(1,0,1)$, and
$\mathbf{R}_3=\frac{a}{2}\,(1,1,0)$, with $a=5.4310$\AA.
The reciprocal lattice vectors are
$\GG_1=\frac{2\pi}{a}\,(-1,1,1)$,
$\GG_2=\frac{2\pi}{a}\,(1,-1,1)$, and
$\GG_3=\frac{2\pi}{a}\,(1,1,-1)$.
The four vectors connecting sublattice $A$ to $B$ are
$\dd_0=\frac{a}{4}\, (1, 1, 1)$,
$\dd_1=\frac{a}{4}\, (1 ,-1, -1)$,
$\dd_2=\frac{a}{4}\, (-1, 1, -1)$, and
$\dd_3=\frac{a}{4}\, (-1, -1, 1)$.

The off-diagonal block matrix, whose dimension depends on how many
orbitals we consider, is given by Eq.~(\ref{eq:off-diagonal_H}) of the
main text with the index $\mu=0,1,2,4$:
\begin{equation}
  \left[H^{\;}_{AB}(\kk)\right]_{\alpha\beta}
  =
  -
  \sum_{\mu=0}^4\;
  t_{\alpha\beta}(\dd_\mu)\;e^{\im\kk\cdot\dd_\mu}
     \;,
\end{equation}
where again we allowed generically for unequal values for the hoppings to
the three nearest neighbors. The vectors $\GG_{1,2,3}$ and
$\dd_{0,1,2,3}$ satisfy
\begin{equation}
  \GG_{i} \cdot \dd_\mu = \frac{2\pi}{4}\;\;(\text{mod}
  \;2\pi)\quad \text{for}\; i=1,2,3\; \text{and}\; \mu=0,1,2,4 \;.
\end{equation}
The Hamiltonian for silicon is therefore not periodic in $\kk$, but
instead is periodic up to the unitary transformation in
Eq.~(\ref{eq:H-gauge}) of the main text [repeated in
  Eq.~(\ref{eq:H-gauge-repeated}) for convenience], with
$\Phi(\GG_i)=2\pi/4$. One can also check that
$-(\GG_{1}+\GG_{2}+\GG_{3}) \cdot \mathbf{d}_\mu =
\frac{2\pi}{4}\;(\text{mod} \;2\pi)$, for $\mu=0,1,2,3$, or
equivalently, $\Phi(-\GG_1-\GG_2-\GG_3)=2\pi/4$.

We then consider the walk in $\kk$-space that was described in the
main text, one that visits, in order, the points
$\kk,\kk+\GG_1,\kk+\GG_1+\GG_2, \kk+\GG_1+\GG_2+\GG_3$ and back to
$\kk$. As explained in the main text, we start at a $\kk$ near but not
at the $\Gamma$ point, to avoid passing through lines of degeneracy.
At the end of the walk, the Hamiltonian returns to $H(\kk)$, but the
eigenvector is rotated by the sequence of unitary operations
\begin{equation}
  \begin{split}
    &U(-\GG_1-\GG_2-\GG_3)\;U(\GG_3)\;U(\GG_2)\;U(\GG_1)
    =
    \\
    &=
    e^{\;\im\frac{1}{2}\left[
        \Phi(-\GG_1-\GG_2-\GG_3)+\Phi(\GG_3)+\Phi(\GG_2)+\Phi(\GG_1)
        \right]\;\sigma_z}
    \\
    &=
    e^{\;\im\pi\;\sigma_z}
    =
    -\oneone
    \;.
  \end{split}
\end{equation}
The wavefunction changes sign upon returning from the walk, which
requires that a $\pi$-flux lines pierce the region enclosed by the
walk. The flux lines required by the argument above are precisely
those described in the main text, obtained via direct calculation of
the Berry curvature in the $sp^3s^*$ model.

\section{Derivation of the effective Hamiltonian at the X points for the conduction band of silicon}

Since electron energies at the X point come in degenerate pairs, the
relevant physics of the two lowest conduction bands (that together are
degenerate along the X-W direction) is to be described by an effective
$2\times 2$ Hamiltonian. This can be done within the framework of
degenerate perturbation theory\cite{ioffe}.

In order to derive such an effective Hamiltonian we first diagonalize
the full $10\times 10$ Hamiltonian of the $sp^3s^*$ model at the X
point to identify pairs of eigenstates corresponding to five
degenerate eigenvalues. Then we expand the Hamiltonian in the newfound
basis to the second order in momentum (around the X point) to get a
matrix of the type
\begin{equation}
	H=\begin{bmatrix}
		\hat{H}_{I_1} & \hat{H}_{I_1I_2} & \dots & \hat{H}_{I_1I_5}
		\\
		\hat{H}_{I_2I_1} & \hat{H}_{I_2} & \dots & \hat{H}_{I_2I_5}
		\\
		\vdots & \vdots & \ddots & \vdots
		\\
		\hat{H}_{I_5I_1} & \hat{H}_{I_5I_2} & \dots & \hat{H}_{I_5}
	\end{bmatrix},
\end{equation}
where $I_\alpha$ label groups of degenerate sublevels and
$H_{I_\alpha}$ are $\mathfrak{D}_\alpha$ by $\mathfrak{D}_\alpha$
matrices with $\mathfrak{D}_\alpha$ being the degeneracy of a given
level group. $\hat{H}_{I_\alpha I_\beta}$ represent matrix elements
between two such groups of levels. In the case of the X point in
silicon all degeneracies are twofold and $\{I_1, I_2, I_3, I_4,
I_5\}=\{(1,2), (3,4), (5,6), (7,8), (9,10)\}$, the enumeration
starting from the lowest valence band. Then within second-order
perturbation theory the effective Hamiltonian describing level group
$I_\alpha$ is given by
\begin{equation}
\hat{H}_{I_\alpha}^{\text{eff}}=\hat{H}_{I_\alpha}-\sum_{\beta\neq \alpha}\frac{\hat{H}_{I_\alpha I_\beta}\hat{H}_{I_\beta I_\alpha}}{E_{I_\alpha}-E_{I_\beta}}.
\end{equation}
Since $I_\alpha$ and $I_\beta$ are groups of indices, we can also
clarify the equation above and expand in a more explicit form using
ordinary level indices $m, n, l$ and the full Hamiltonian $H$ as
\begin{equation}
\left(\hat{H}_{I_\alpha}^{\text{eff}}\right)_{m,n}=H_{m,n}-\sum_{\beta\neq \alpha}\frac{1}{E_{I_\alpha}-E_{I_\beta}}\sum_{l\in I_\beta}H_{m,l}H_{l,n},\quad m,n\in I_\alpha.
\end{equation}

Performing this procedure for silicon numerically, we obtain the
effective Hamiltonian for bands 5 and 6 that was presented in the main
text,
\begin{equation}
	H_{(5,6)}^{\text{eff}}=
	\begin{bmatrix}
		1.63 + 0.02p_x^2 & 0.29p_yp_z+0.02ip_x
		\\
		0.29p_yp_z-0.02ip_x & 1.63 + 0.02p_x^2
	\end{bmatrix}.	
\end{equation}

\end{widetext}
%








\end{document}